\documentclass[12pt]{article}

\usepackage{scicite}

\usepackage{times,amsmath}
\usepackage{color}

\usepackage{graphicx}

\newenvironment{sciabstract}{%
\begin{quote} \bf}
{\end{quote}}

\newcounter{lastnote}

\title{Humans display a reduced set of consistent behavioral phenotypes in dyadic games}

\author
{Julia Poncela-Casasnovas,$^{1}$ Mario Guti\'errez-Roig, $^{2}$ \\
Carlos Gracia-L\'azaro, $^{3}$ Julian Vicens, $^{1,4}$ \\
Jes\'us G\'omez-Garde\~nes, $^{3,5}$ Josep Perell\'o, $^{2}$ Yamir Moreno, $^{3,6,7}$ \\
Jordi Duch, $^{1}$ Angel S\'anchez $^{3,8,9 \ast}$\\
\\
\normalsize{$^{1}$Departament d'Enginyeria Inform\`atica i Matem\`atiques, Universitat Rovira i Virgili,}\\
\normalsize{43007, Tarragona, Spain}\\
\normalsize{$^{2}$Departament de F\'isica Fonamental, Universitat de Barcelona,}\\
\normalsize{08028 Barcelona, Spain}\\
\normalsize{$^{3}$Institute for Biocomputation and Physics of Complex Systems (BIFI), University of Zaragoza,}\\
\normalsize{50018 Zaragoza, Spain}\\
\normalsize{$^{4}$Applied Research Group in Education and Technology, Universitat Rovira i Virgili,}\\
\normalsize{43007, Tarragona, Spain}\\
\normalsize{$^{5}$Department of Condensed Matter Physics, University of Zaragoza,}\\
\normalsize{50009 Zaragoza, Spain}\\
\normalsize{$^{6}$Department of Theoretical Physics, University of Zaragoza,}\\
\normalsize{50009 Zaragoza, Spain}\\
\normalsize{$^{7}$Complex Networks and Systems Lagrange Lab, Institute for Scientific Interchange,}\\
\normalsize{Turin, Italy}\\
\normalsize{$^{8}$Grupo Interdisciplinar de Sistemas Complejos, Departamento de Matem\'aticas, }\\
\normalsize{Universidad Carlos III de Madrid, 28911 Legan\'es, Madrid, Spain}\\
\normalsize{$^{9}$Institute UC3M-BS of Financial Big Data, Universidad Carlos III de Madrid,}\\
\normalsize{28903 Getafe, Spain}\\
\\
\normalsize{$^\ast$To whom correspondence should be addressed; E-mail: anxo@math.uc3m.es}
}

\date{}

\begin{document} 


\baselineskip24pt


\maketitle


\begin{sciabstract}
Socially relevant situations that involve strategic interactions are widespread among animals and humans alike. To study these situations, theoretical and experimental works have adopted a game-theoretical perspective, which has allowed to obtain valuable insights about human behavior. However, most of the results reported so far have been obtained from a population perspective and considered one specific conflicting situation at a time. This makes it difficult to extract conclusions about the consistency of individuals' behavior when facing different situations, and more importantly, to define a comprehensive classification of the strategies underlying the observed behaviors. Here, we present the results of a lab-in-the-field experiment in which subjects face four different dyadic games, with the aim of establishing general behavioral rules dictating individuals' actions. By analyzing our data with an unsupervised clustering algorithm, we find that all the subjects conform, with a large degree of consistency, to a limited number of behavioral phenotypes (Envious, Optimist, Pessimist, and Trustful), with only a small fraction of undefined subjects. We also discuss the possible connections to existing interpretations based on a priori theoretical approaches. Our findings provide a relevant contribution to the experimental and theoretical efforts towards the identification of basic behavioral phenotypes in a wider set of contexts without aprioristic assumptions regarding the rules or strategies behind actions. From this perspective, our work contributes to a fact-based approach to the study of human behavior in strategic situations, that could be applied to simulating societies, policy-making scenario building and even for a variety of business applications.  
\end{sciabstract}

\section*{Introduction}

Many situations in life entail social interactions where the parties involved behave strategically: That is, they take into consideration the anticipated responses of actors who might otherwise have an impact on an outcome of interest. Examples of such interactions include social dilemmas, where individuals face a conflict between self and collective interests, which can also be seen as a conflict between rational and irrational decisions \cite{dawes:1980,kollock:1998,vanlange:2013}, as well as coordination games, where all parties are rewarded for making mutually consistent decisions \cite{skyrms:2003}. These and related scenarios are commonly studied in economics, psychology, political science and sociology, typically using a game theoretic framework to understand how decision makers approach conflict and cooperation under highly simplified conditions \cite{sigmund:2010,gintis:2009,myerson:1991}.

Extensive work has shown that, when exposed to the constraints introduced in game theory designs, people are often not ``rational'' in the sense that they do not pursue exclusively self-interested objectives \cite{camerer:2003,kagel:1997} . This is especially clear in the case of Prisoner's Dilemma (PD) games, where rational choice theory predicts that players will always defect but empirical observation shows that cooperation oftentimes occurs, even in ``one-shot''  games where there is no expectation of future interaction among the parties involved \cite{camerer:2003,ledyard:1997} . These findings beg the question as to why players sometimes choose to cooperate despite incentives not to do so. Are such choices a function of a person's identity and therefore consistent across different strategic settings? Do individuals draw from a small repertoire of responses, and if so, what are the conditions that lead them to choose one strategy over another? 

Here, we attempt to shed light on these questions by focusing on a wide class of simple dyadic games that capture two important features of social interaction, namely the temptation to free-ride and the risk associated with cooperation \cite{camerer:2003,rapoport:1966,macy:2002}. All games are two-person, two-action games in which participants decide simultaneously which of the two actions they will take. Following previous literature, we classify participants' set of choices as either cooperation, which we define as a choice that promotes the general interest, or defection, a choice that serves an actor's self interest at the expense of others. 
The games utilized in our study include Prisoner's Dilemma \cite{rapoport:1965,axelrod:1981}, the Stag Hunt \cite{skyrms:2003}, and the Hawk-Dove  \cite{smith:1982} or Snowdrift \cite{sugden:2005} games. Stag Hunt (SH) is a coordination game in which there is a risk in choosing the best possible option for both players: cooperating when the other party poses serious consequences for the cooperator, while the defector faces less extreme costs for non-cooperation \cite{cooper:1998}. Hawk-Dove (SG) is an anti-coordination game where one is tempted to defect, but participants face the highest penalties if both player defect \cite{bramoulle:2007}. In Prisoner's Dilemma (PD) games, both tensions are present: when a player defects, the counterpart faces the worst possible situation, whereas the defector benefits more than cooperating. We also consider the Harmony Game (HG), where the best individual and collective options coincide and therefore there should be no tensions \cite{licht:1999}.

Several theoretical perspectives have sought to explain the seemingly irrational behavior of actors during conflict and cooperation games. Perhaps most prominent among them is the theory of social value orientations \cite{VanLange:2000,rusbult:2003,balliet:2009}, which focuses on how individuals divide resources between the self and others. This research avenue has found that individuals tend to fall into certain categories such as individualistic (think only about themselves), competitive (want to maximize theirs and others' payoffs), cooperative (attempt to maximize everyone's outcome), and altruistic (sacrifice their own benefits to help others). Relatedly, social preferences theory posits that people's utility functions often extend beyond their own material payoff and may include considerations of aggregate welfare or inequity aversion \cite{fehr:1999} . Whereas theories of social orientation and social preferences assume intrinsic value differences between individuals, cognitive hierarchy theory instead assumes that players make choices based on their predictions about the likely actions of other players, and as such, the true differences between individuals comes not from values but rather from depth of strategic thought \cite{camerer:2004} .

One way to arbitrate between existing theoretical paradigms is to use within-subject experiments, where participants are exposed to a wide variety of situations requiring strategic action. If individuals exhibit a similar logic (and corresponding behavior) in different experimental settings, this would provide a more robust empirical case for theories that argue strategic action stems from intrinsic values or social orientation. By contrast, if participants' strategic behavior depends on the incentive structure afforded by the social context, such findings would pose a direct challenge to the idea that social values drive strategic choices. 

We therefore contribute to the literature on decision making in three important ways. First, we expose the same participants to multiple games with different incentive structures in order to assess the extent to which strategies stem from stable characteristics of an individual. Second, we depart from existing paradigms by not starting from an a priori classification to analyze our experimental data. For instance, empirical studies have typically utilized classifications schemes that were first derived from theory, making it difficult to determine whether such classifications are the best fit for the available data. We address this issue by using an unsupervised, robust classification algorithm to identify the full set of ``strategic phenotypes''  that constitute the repertoire of choices among individuals in our sample. Finally, we advance research that documents the profiles of cooperative phenotypes \cite{peysakhovich:2014} by expanding the range of human behaviors that may exhibit similar types of classification. In focusing on both cooperation and defection, this approach allows us to make contributions towards a taxonomy of human behaviors \cite{kirman:1992,blanco:2011}.

\section*{Lab-in-the-field experiment}

We recruited 541 subjects of different ages, education level and social status during a fair in
Barcelona \cite{sagarra:2015} (see Materials and
Methods). The experiment consisted of multiple rounds, in which participants were randomly assigned partners and assigned randomly chosen payoff values, 
allowing us to study the behavior of the same subject in a variety of dyadic games including PD, SH, SG and
HG, with different payoffs. In order to incentivize the experimental subjects' decisions with real material (economic)
consequences, they were informed that they would receive lottery tickets proportionally (one ticket per each 40 points; the modal number of tickets earned was 2) to the payoff they accumulated
during the rounds of dyadic games they played. The prize in the corresponding lottery was 4 coupons redeemable at
participating neighboring stores, worth 50 euros each.
The payoff matrices shown to the participants had the following form (rows are participant's strategies while columns are
the opponent's ones):
\begin{eqnarray}
\bordermatrix{
 & C & D \cr
C & R & S \cr
D & T & P \cr}
\label{tab:payoffMatrix} 
\end{eqnarray}
Actions $C$ and $D$ were coded as two randomly chosen colors in the experiment to avoid framing effects. $R$ and $P$ were always set to $R=10$ and $P=5$ whereas $T$ and $S$ took values $T \in [5,15]$ and $S \in [0,10]$. In this way, the $(T,S)$-plane can be divided into four quadrants- , each one corresponding to a different game depending on the relative order of the payoffs: HG ($S>P,\, R>T$), SG ($T>R>S>P$), SH ($R>T>P>S$) and PD ($T>R>P>S$). Matrices were generated with equal probability for each point in the $(T,S)$-plane, which was discretized as a lattice of $11 \times 11$ sites. Points in the boundaries between games, at the boundary of our game space, or in its center do not correspond to the four basic games to which we refer. However, we kept those points to add generality to our exploration and, in any event, we made sure in the analysis that the results did not change if we removed those special games (see below). For reference, see Fig.~1 (middle) for the Nash (symmetric) equilibrium structure of each one of these games.

\section*{Population level behavior}

The average level of cooperation aggregated over all games and subjects is $ \langle C \rangle=0.49\pm 0.01$, where the error corresponds to a 95\% Confidence Interval (we apply this rule to the rest of our results, unless otherwise specified). This is in very good agreement with the theoretically expected value, $ \langle C \rangle^{theo}=0.5$, calculated by averaging over all the symmetric Nash equilibria for the $(T,S)$ values analyzed. However, the aggregate cooperation heatmap looks very different from what would be obtained by simulating a population of players on a well-mixed scenario (compare right and central panels in Fig.~1).

On the ohter hand, the experimental levels of cooperation per game (excluding the boundaries between them, so the points strictly correspond to one of the four games) are: $\langle C \rangle_{PD}=0.29\pm 0.02$ ($\langle C \rangle^{theo}_{PD}=0$), $\langle C \rangle_{SG}= 0.40\pm 0.02$ ($\langle C \rangle^{theo}_{SG}=0.5$), $\langle C \rangle_{SH}= 0.46\pm 0.02$ ($\langle C \rangle^{theo}_{SH}=0.5$), and $\langle C \rangle_{HG}= 0.80\pm 0.02$ ($\langle C \rangle^{theo}_{HG}=1$). The values are considerably different from the theoretical ones in all cases, particularly for the PD and HG.

\begin{figure}[!th]   
\begin{center}$ 
\begin{array}{cc}
\includegraphics[width=0.95\textwidth,angle=0]{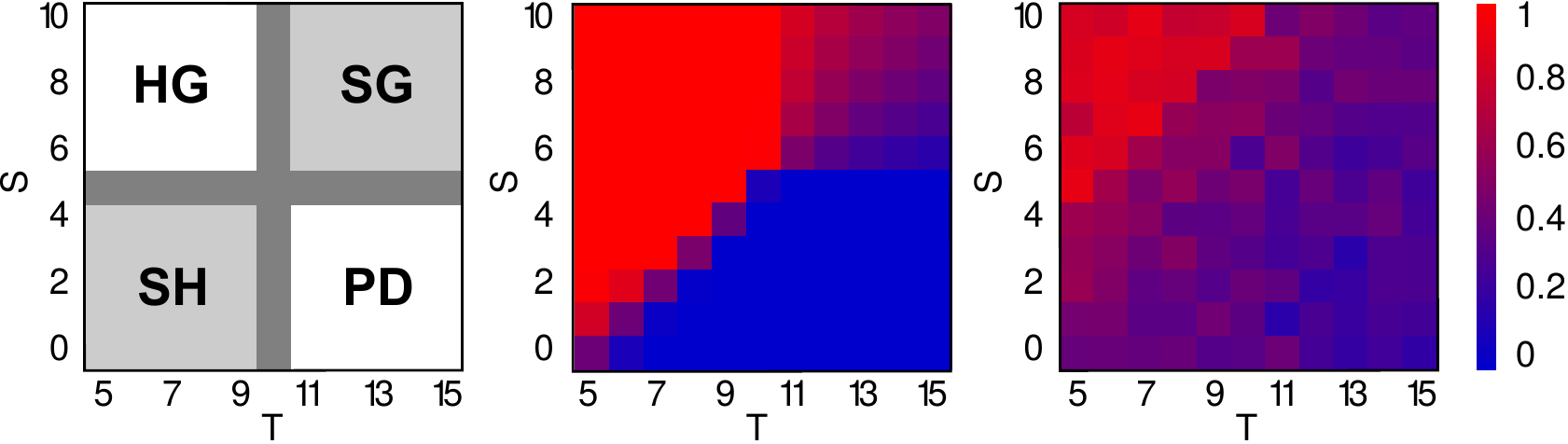}
\end{array}$
\end{center}
\caption{Schema with labels to help identify each one of the games in the quadrants of the $(T,S)$-plane (left), along with the theoretical equilibria (center) and average empirical cooperation heatmaps from the $8,366$ game actions of the $541$ subjects (right), in each cell of the $(T,S)$-plane. The Nash Equilibria for each game are (center): Prisoner's Dilemma (PD) and Harmony Game (HG) have one equilibrium, given by the pure strategy $D$ and $C$ respectively. Snowdrift Game (SG) has a stable mixed equilibrium containing both cooperators and defectors, in a proportion that depends on the specific payoffs considered. SH is a coordination game displaying two pure-strategy stable equilibria, whose bases of attraction are separated by an unstable one, again depending on the particular payoffs of the game \cite{harsanyi:1988,gintis:2009,sigmund:2010}. The fraction of cooperation is color coded from red (full cooperation), to blue (full defection).}\label{fig:cooperation_heatmap}
\end{figure}

\begin{figure}[!th]   
\begin{center}$ 
\begin{array}{cc}
\includegraphics[width=0.95\textwidth,angle=0]{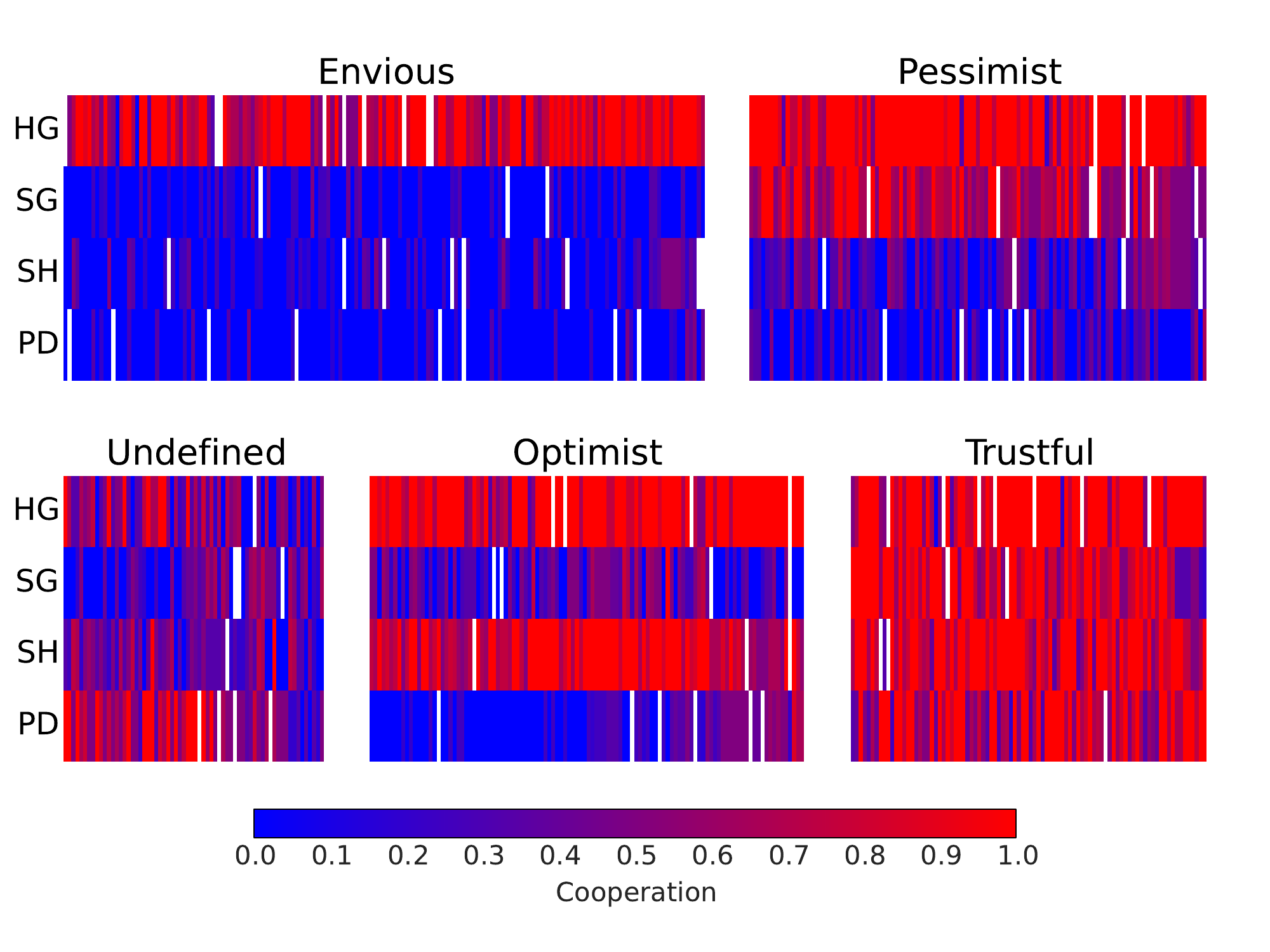}
\end{array}$
\end{center}
\caption{Results from the $K$-means clustering algorithm. For every one of the five clusters, each column represents a
player belonging to that cluster, while the four rows are the four average cooperation values associated to her (from top
to bottom: cooperation in HG, SG, SH and PD games). We color-coded the average level of cooperation for each player in
each game from blue (0.0) to red (1.0), while the lack of value in a particular game for a particular player is coded in
white. Cluster sizes: Envious, $N=161$ (30\%); Pessimist, $N=113$ (21\%); Undefined, $N=66$(12\%); Optimist $N=110$ (20\%); and Trustful $N=90$ (17\%).}\label{fig:clustering}
\end{figure}

\begin{figure}[!th]   
\begin{center}
\includegraphics[width=0.95\textwidth]{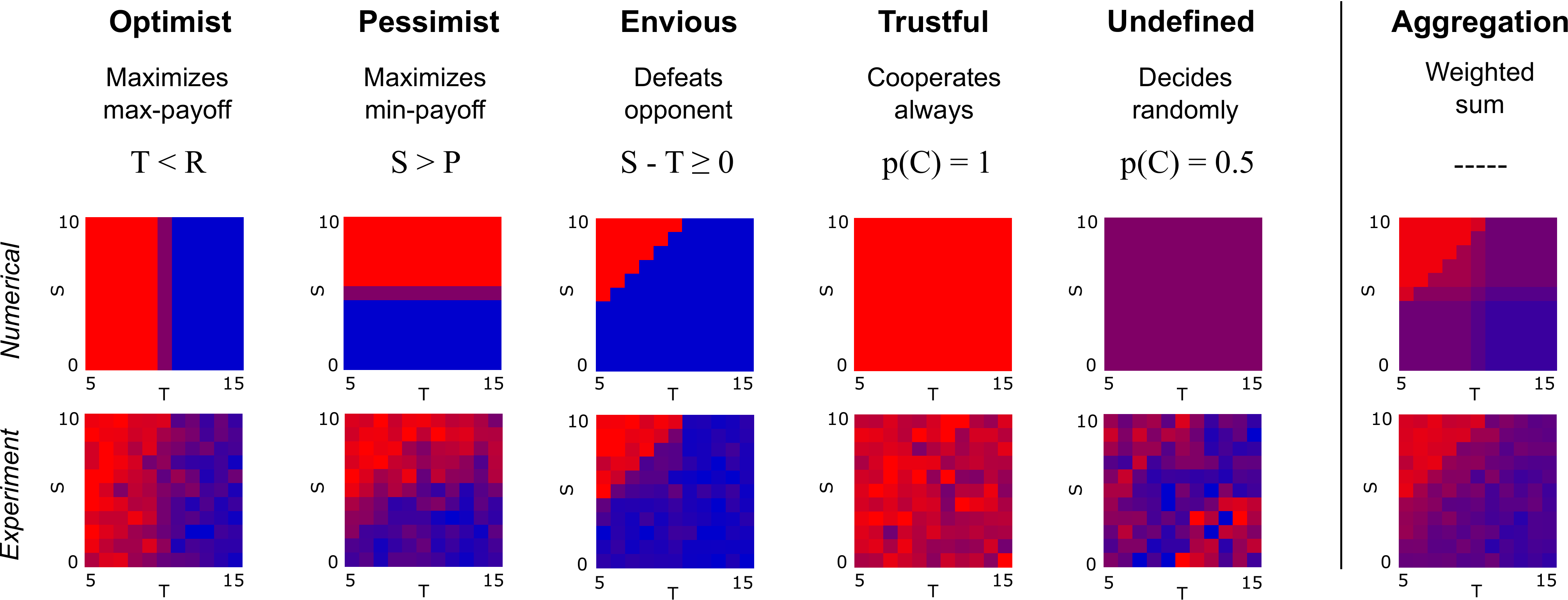}
\end{center}
\caption{Summary results of the different phenotypes (Optimist, Pessimist, Envious, Trustful and Undefined) determined by the $K$-means clustering algorithm, plus the aggregation of all phenotypes. For each phenotype (column), we show the word description of the behavioral rule, and the corresponding inferred behavior in the whole $(T,S)$-plane (labeled as Numerical). The fraction of cooperation is color coded from red (full cooperation), to blue (full defection). The last row (labeled as Experiment), shows the average cooperation, aggregating all the decisions taken by the subjects classified in each cluster. The fraction of each phenotype is: 20\% Optimist, 21\% Pessimist, 30\% Envious, 17\% Trustful, and 12\% Undefined. The very last column shows the aggregated heatmaps of cooperation both for the simulations (assuming that each individual plays using one and only one of the behavioral rules, and respecting the relative fractions of each phenotype in the population found by the algorithm), and the experimental results. Note the good agreement between aggregated experimental and aggregated numerical heatmaps (the discrepancy heatmap between them is shown in SI Appendix, Sec.\ S.4.11). We report that, the average difference across the entire $(T,S)$ plane between the experiment and the phenotype aggregation is of $1.39$ SD units, which represents a value inside the standard 95\% Confidence Interval.  While for any given phenotype, this difference averaged over all $(T,S)$-plane is smaller than $2.14$ SD units.
}
\label{fig:panels_cooperation_TS_by_clustering}
\end{figure}

\section*{Emergence of phenotypes}

After looking at the behavior at the population level, we focus on the analysis of the decisions at the individual level
\cite{kirman:1992}. Our goal is to asses whether individuals behave in highly idiosyncratic manners or if, on the contrary,
there are only a few `phenotypes' in which all our experimental subjects can be classified. To this aim, we characterize each subject with a 4-dimensional vector where each dimension represents her average level of cooperation in each one of the four quadrants in the $(T,S)$-plane. Then, we apply an unsupervised clustering procedure, the $K$-means clustering algorithm \cite{macqueen:1967}, to group those individuals that have similar behaviors, i.e. the values in their vectors are similar. This algorithm (see SI Appendix, Sec.\
S.4.7) takes as input the number of clusters $k$ to be found, and it groups the data in such a way that it both minimizes
the dispersion within clusters and maximizes the distance among centroids of different clusters. We found that $k=5$
clusters is the optimal number of groups according to the Davies-Bouldin index \cite{davies_bouldin:1979} (see SI Appendix, Sec.\ S.4.8), which does not assume beforehand any specific number of types of behaviors.

The results of the clustering analysis are presented in Fig.~2., and they show that there is a group that mostly cooperates in the HG, a second group that cooperates in both HG and SG, and a
third one that cooperates in both HG and SH. Players in the fourth group cooperate in all games, and finally, we find a
small group who seem to randomly cooperate with probability approximately $0.5$ almost everywhere.

In order to obtain a better understanding of the behavior of these five groups, we represent the different types of behavior in a heatmap (Fig.~3) to extract characteristic behavioral rules.  In this respect, it is important to note that Fig.~3 provides a complementary
view of the clustering results: our clustering analysis was carried out attending only to the aggregate cooperation level per quadrant, i.e., to four numbers or coordinates per subject, while this plot shows {\it for every point in the space of games} the average number of times the players in each group cooperated.

The cooperation heatmaps in Fig.~3 show that there are indeed common characteristics to subjects
classified in the same group even when looking at every point of the (S,T) plane. 
The first two columns in Fig.~3 display consistently different behaviors in coordination and anti-coordination games although they both act as prescribed by the Nash equlibrium in PD and HG. Interestingly, both groups are amenable to a simple interpretation that links them to well-known behaviors in economic theory. Thus, the first phenotype ($N=110$, or $20\%$ of the population) defects wherever $T > R$, {i.e.}, they cooperate in the HG and in the SH, and defect otherwise.
By using this strategy, these subjects aim to obtain the maximum  payoff without taking into account the likelihood that their counterpart will allow them to get it, in agreement with a \textit{maximax} behavior \cite{colman:1995}. Accordingly, we
call this first phenotype  `Optimists'. 
Conversely, we label subjects in the second phenotype `Pessimists' (group size $N=113$, or 21\% of the population), as they use a \textit{maximin} principle \cite{vonneumann:1944} to choose their actions, cooperating only when $S > P$, {i.e.}, in the HG and in the SG, in order to ensure a best worst-case scenario. The behaviors of these two phenotypes which, as discussed by Colman \cite{colman:1995}, can hardly be considered rational, are also associated to different degrees of risk aversion, a question that will be addressed below.

Regarding the third column in Fig.~3, it is apparent from the plots that individuals in this phenotype (size $N=161$, or 30\% of the population) cooperate in the upper triangle of the HG exclusively, i.e., wherever $(S-T) \geq 0$. As was the case with 'Optimists' and 'Pessimists', this third behavior is far from being rational in a self-centered sense, in so far as players forsake the possibility of achieving the maximum payoff by playing the only Nash equilibrium in the HG. In turn, these
subjects seem to behave as driven by envy, status-seeking consideration, or lack of trust. By choosing $D$ when $S>P$ and $R>T$ these players prevent their counterparts from receiving more payoff than themselves even when, by doing so, they diminish their own potential payoff. The fact that competitiveness overcomes rationality as players basically attempt to ensure they receive more payoff than their opponents suggests an interpretation of the game as an Assurance game\cite{vanlange:2013} and, accordingly, we have dubbed this phenotype `Envious'.

The fourth phenotype (fourth column in Fig.~3) includes those players who cooperate almost in every round and almost in every site of the $(T,S)$-plane (size $N=90$, or 17\% of the population). In this case, and opposite to the previous one, we believe that these players' behavior can be associated with trust on partners behaving in a cooperative manner. Another way of looking at trust in this context is in terms of expectations, as it has been shown that expectation of cooperation enhances cooperation in the PD \cite{ng:2015}. In any event,
explaining the roots of this type of cooperative behavior in a unique manner seems to be a difficult task, and in fact alternative explanations of cooperation on the PD involving normalized measures of greed and fear \cite{ahn:2001} or up to five simultaneous factors  \cite{engel:2012} have been advanced too. Lacking an unambiguous motivation of the observed actions of the subjects in this group, we find the name `Trustful' to be an appropriate one to refer to this phenotype. Lastly, the unsupervised algorithm found a small fifth group of players (size $N=66$, or 12\% of the population) who cooperate in an approximately random manner, with probability $0.5$, in any situation. For lack of a better insight on their behavior, we will refer to this minority as `Undefined' hereafter.

Remarkably, three of the phenotypes reported here (Optimist, Pessimist and Trustful) have a very similar size. On the other hand, the largest one is the Envious phenotype, including almost a third of the participants, whereas the Undefined group, that can not be considered a {\em bona fide} phenotype in so far as we have not found any interpretation of the corresponding subjects' actions, is considerably smaller than all the others. In agreement with abundant experimental evidence, we have not found any purely rational phenotype: the strategies used by the four relevant groups are, to different extents, quite far from self-centered rationality. Note that ours is an across-game characterization, which does not exclude the possibility of subjects taking rational, purely self-regarding decisions when restricted to one specific game (see SI Appendix, Sec.\ S.4.5).

Finally, and to shed more light on the phenotypes found above, we estimate an indirect measure of their risk aversion. To do this, we consider the number of cooperative actions in the SG together with the number of defective actions in the SH (over the total sum of actions in both quadrants for a given player, see SI Appendix, Sec.\ S.4.5). While Envious, Trustful, and Undefined players exhibit intermediate levels of risk aversion ($0.52$, $0.52$ and $0.54$, respectively), Pessimists exhibit a significantly higher value ($0.73$), consistent with their fear of facing the worst possible outcome and their choice of the best worst-case scenario. In contrast, the Optimist phenotype shows a very low risk aversion ($0.32$), in agreement with the fact that they aim to obtaining the maximum possible payoff, taking the risk that their counterpart do not work with them towards that goal.

\section*{Robustness of phenotypes}

We have carefully checked that our $K$-means clustering results are robust. Lacking the `ground truth' behind our data in
terms of different types of individual behaviors, the significance and robustness of our clustering analysis must be
tested by checking its dependence on the dataset itself. We studied this issue in several complementary manners. First,
we applied the same algorithm to a randomized version of our dataset (preserving the total number of cooperative actions
in the population, but destroying any correlation among the actions of any given subject), showing no significant
clustering structure at all (see  SI Appendix, Sec.\ S.4.7 for details). 

Second, we ran the $K$-means clustering algorithm on portions of the original data with
the so-called `leave-$p$-out' procedure \cite{kohavi:1995}. This test showed that the optimum 5-cluster scheme found is
robust even when randomly excluding up to $55\%$ of the players and their actions (see SI Appendix, Sec.\ S.4.7 for details). Moreover, we repeated the whole analysis discarding the first two choices made by every player, to account for excessive noise due to initial lack of experience, and the results show even more clearly the same optimum at five phenotypes. See SI Appendix, Sec.\ S.4.7 for a complete discussion.

Third, we tested the consistency among cluster structures found in different runs of the same algorithm for a fixed
number of clusters, that is to say, how likely it is that the particular composition of individuals in the cluster scheme
from one realization of the algorithm is correlated with the composition from that of a different realization. To ascertain this, we
computed the Normalized Mutual Information Score, $MI$,  \cite{MacKay:2003} (see SI Appendix, Sec.\ S.4.9 for formal definition)
knowing that the comparison of two runs with exactly the same clustering composition would give a value
$MI=1$ (perfect correlation), and $MI=0$ would correspond to a total lack of correlation between them. We ran our
$K$-means clustering algorithm $2,000$ times for the optimum $k=5$ clusters and we paired the clustering schemes for comparison, obtaining
an average Normalized Mutual Information score of $MI=0.97$  ($SD:0.03$). To put these numbers in perspective, the
same score for the pair-wise comparison of results from $2,000$ realizations of the algorithm on the randomized version
of the data gives $MI=0.59$ ($SD:0.18$), see SI Appendix, Sec.\ S.4.9 for more details. 

All the tests presented above provide strong support for our classification in terms of phenotypes. However, we also searched
for possible dependencies of the phenotype classification on the age and gender distributions for each group (see SI Appendix, Sec.\ S.4.10), and we found no significant differences among them, which hints towards a classification of
behaviors (phenotypes) beyond demographic explanations.

\section*{Discussion and conclusions}
\label{sec:conclusions}

We have presented the results of a lab-in-the-field experiment designed to identify `phenotypes', following the terminology fittingly introduced by Peysakhovich et al. \cite{peysakhovich:2014}. 
Our results suggest that the individual behaviors of the subjects in our population can be described by a small set of phenotypes: Envious, Optimist, Pessimist, Trustful and a small group of individuals referred to as 'Undefined' who play an unknow strategy.
The relevance of this repertoire of phenotypes arises from the fact that it has been obtained from experiments in which subjects played a wide variety of dyadic games, through an unsupervised procedure, the $K$-means clustering algorithm, and that it is a very robust classification. With this technique, we can go beyond correlations and
assign specific individuals to specific phenotypes, instead of looking at (aggregate) population data. In this respect, the tri-modal distributions of the joint cooperation probability found by Capraro et al. \cite{capraro:2014} show much resemblance with our findings, and, while a direct comparison is not possible because they correspond to aggregate data, they point in the direction of a similar phenotype classification. In addition, our results contribute to the currently available evidence that people are heterogeneous, by quantifying the degree of heterogeneity{{color{red},} in terms of both the number of types and their relative frequency, in a specific (but broad) suite of games.
 
While the robustness of our agnostic identification of phenotypes makes us confident of the relevance of the behavioral classification, and our interpretation of it is clear and plausible, it is not the only possible one. It is important to point out that connections can also be drawn to earlier attempts to classify individual behaviors. As we have mentioned above, one theory that may also shed light on our classification is that of social value orientation \cite{VanLange:2000, rusbult:2003, balliet:2009}. Thus, the Envious type may be related to the competitive behavior found in that context (although in our observation envious people just aim at making more profit than their competitors, not necessarily minimizing their competitors' profit); Optimists could be cooperative, and Trustful seem very close to altruistic. 
As for the Pessimist phenotype, we have not been able to draw a clear relationship to the types most commonly found among social value orientations, but in any event the similarities between the two classifications is appealing and suggests an interesting line for further research. Another alternative view on our findings arises from social preferences theory \cite{fehr:1999}, where, for instance, envy can be understood as the case in which inequality that is advantageous to self yields a positive contribution to one's utility \cite{bolton:2000,charness:2002,cabrales:2010,cabrales:2010b}. Altruists can be viewed as subjects with concerns for social welfare \cite{charness:2002}, whereas for the other phenotypes, we find it difficult to understand them in this framework and, in fact, optimists and pessimists do not seem to care about their partner's outcome. However, other interpretations may apply to these cases: Optimists could indeed be players strongly influenced by payoff dominance {\em «a la} Harsanyi and Selten \cite{harsanyi:1988}, in the sense that these players would choose strategies associated with the best possible payoff for both. Yet another view on this phenotype is that of team reasoning \cite{bacharach:1999,sugden:1993,sugden:2011}, namely individuals whose strategies  maximize the collective payoff of the player pair   if such a strategy profile is unique. Interestingly, proposals such as the cognitive hierarchy theory \cite{camerer:2004, colman:2014}, and the level-k theory \cite{stahl:1994, stahl:1995} do not seem to fit our results, in so far as the best response to the undefined phenotype, which would be the zeroth level of behavior, does not match any of our behavioral classes. 

Our results open the door to making relevant advances in a number of directions. For instance, they point to the independence of 
the phenotypic classification on age and gender. While the lack
of gender dependence may not be surprising, it would be really astonishing that small children would exhibit behaviors with similar classifications in view of the body of experimental evidence about their differences with adults \cite{fehr:2008,house:2013,charness:2009,sutter:2007,benenson:2007,gutierrezroig:2014}, and further research is needed to assess this issue in detail. In fact, as discussed also by Peysakhovich
et al. \cite{peysakhovich:2014}, our research does not illuminate whether the different phenotypes are born, made, or something in between, and understanding their origin would then be 
a far-reaching result. 

We believe that applying an approach similar to ours to results about the cooperative phenotype \cite{yamagishi:2013,peysakhovich:2014,capraro:2014} and, even better, to carry out experiments with an ample suite of games, as well as a detailed questionnaire \cite{exadaktylos:2013} is key in future research. In this regard, it has to be noted that the relationship between our automatically identified phenotypes and theories of economic behavior yields predictions about other games: indeed, envy and expectations about the future and about other players will dictate certain behaviors in many other situations. Therefore, our classification here can be tested and refined by looking for phenotypes arising in different contexts. This could be complemented with a comparison of our unsupervised algorithm with the parametric modeling approach in  \cite{cabrales:2010} or even implementing flexible specifications to social preferences\cite{fehr:1999,bolton:2000,charness:2002}  or social value orientation \cite{VanLange:2000, rusbult:2003, balliet:2009} to improve the understanding of our behavioral phenotypes.

Finally, our results have also implications in policy making and real-life economic interactions. For instance, there is a large group of individuals, the Envious ones (about a third of the population), that in situations such as the Harmony game, fail to cooperate when they are at risk of being left with lower payoff than the counterpart. This points to the difficulty 
of making people understand  
when they face a non-dilemmatic, win-win, situation, and that effort must be put to make this very clear.  Another interesting sub-population is that of Pessimist and Optimist phenotypes, which together amount to approximately half of the population. These people exhibit large or small risk aversion, respectively, and use an ego-centered approach in their daily lives, thus ignoring that others can improve or harm their expected benefit with highly undesirable consequences. A final example of the hints provided by our results is the existence of an unpredictable fraction of the population (Undefined) that, even being small, can have a strong influence in social interactions because their noisy behavior could lead people with more clear heuristics to mimic their erratic actions. On the other hand, the classification in terms of phenotypes (particularly if, as we show here, comprises only a few different types) can be very useful for firms, companies or banks interacting with people: it could be used to evaluate customers or potential ones, or even employees for managerial purposes, allowing for a more efficient handling of the human resources in large organizations. Such approach is indeed also very valuable in the emergent deliberative democracy and open government practices around the globe (including the behavioral Insights Team \cite{bituk} of the UK Government, its recently established counterpart at the White House, or the World Health Organization \cite{who}). Research following the lines presented here can lead to many innovations in these contexts.

\section*{Materials and Methods}
The experiment was conducted as a lab-in-the-field one, i.e., to avoid restricting ourselves to the typical samples of
university undergraduate students, we took our lab to a festival in Barcelona and recruited subjects from the general audience there \cite{sagarra:2015}. This setup allows, at the very least, to obtain results from a very wide age range, as was the case in a previous study where it was found that teenagers behave differently \cite{gutierrezroig:2014}. All participants in the experiment signed an informed consent to participate. In agreement with the Spanish Law for Personal Data Protection, no association was ever made between their real names and the results. This procedure was checked and approved by the Viceprovost of Research of Universidad Carlos III de Madrid, the institution funding the experiment. 

In order to cover equally the four dyadic games in our experiments, we discretized the $(T,S)$-plane as a lattice of
$11 \times 11$ sites.
Each player was equipped with a tablet running the application of the experiment (see SI Appendix, Sec.\ S.1 for technical details and
Sec.\ S.2 for the experiment protocol). The participants were shown a brief tutorial in the tablet (see the translation
of the tutorial on SI Appendix, Sec.\ S.3), but were not instructed in any particular way nor with any particular goal in mind.
They were informed that they had to make decisions in different conditions and against different opponents in every
round. They were not informed about how many rounds of the game they were going to play. 
Due to practical limitations, we could only host around 25 players simultaneously, so the experiment was conducted in
several sessions over a period of two days. In every session, all individuals played a different, randomly picked number
of rounds between 13 and 18. 
In each round of a session each participant was randomly assigned a different opponent and a payoff matrix corresponding
to a different $(T,S)$ point among our $11\times 11$ different games. Couples and payoff matrices were randomized in each
new round, and players did not know the identity of their opponents. In case there was an odd number of players, or a
given player was non-responsive, the experimental software took over and made the game decision for her, labeling its
corresponding data accordingly to discard it in the analysis ($143$ actions). When the action was actually carried out by the software, the stipulation was that it repeated the previous choice of C or D with an 80\% probability. In the three cases that a session had an odd number of participants, it has to be noted that no subjects played all the time against the software, as the assignment of partner was randomized for every round. The total number of participants in our
experiment was $541$, adding up to a total of $8,366$ game decisions collected, with an average number of actions per
$(T,S)$ value of $69.1$ (see also SI Appendix, Sec.\ S.4.3).

\newpage

\section*{Supplementary Material}

\section*{S.1.\ Technical implementation of the experiment}

\label{sec:implementation_exp}
To conduct the experiment and collect the data we implemented a local network architecture (see Fig.~\ref{fig:system_architecture}) which consisted of $25$ mobile devices (tablets), a router, and a laptop running a web server and a database server. The system was designed to allow playing synchronized sessions, to collect and store user data safely, and to control in real time the experiment while the users were playing against each other.

The game was accessible through a web application specifically designed for tablets. All the interactions that users made through the game interface were immediatelly sent to the server through a client API -no data was stored in the tablets-. The server also provided a server API to control and monitor the status of each experiment session.

The software of the experiment was developed using Django framework and Javascript. Both APIs were implemented using RESTful services and JSON objects for the exchange of data between server and clients, which was stored in a MySQL database.

\section*{S.2.\ Running the experiment}
\label{sec:execution_exp}

The experiment was carried out during the game festival (Festival del Joc) DAU Barcelona {\bf http://lameva.barcelona.cat/daubarcelona}, in December 2014, over a period of two days. We collected data from $541$ subjects in total, who were recruited by our team among the game fair attendees. Due to space limitations, the experiment took place in multiple sessions over those two days, in groups of $15-25$ people. The average age among our $541$ subjects was $31.3$ (SD=$14.3$) (see Fig.~\ref{fig:age_distrib} for the age distribution of the population), with $64.5\%$ males and $35.5\%$ females.

Each person was given a tablet to play the game using the tablet's browser. Before the actual experiment started, the subjects were shown a tutorial in their tablets, to learn (i) the basic rules of how to play the game, (ii) an explanation about the meaning of the payoff matrix and their possible choices, and (iii) a couple of examples of game rounds equivalent to the ones they would face during the actual game. Also, some of our team members were walking around the room answering questions from the subjects during the tutorial period (but not during the actual game). Nonetheless, we did not instruct them to play in any particular way nor with any one particular goal in mind. In Fig.~\ref{fig:tutorial_1} we show the tutorial screens. After a player had read the tutorial, she pressed a button to indicate the system that she was ready to start playing. Once everyone was ready, the game administrator started the game.

Each game session was carried out for a random number of rounds, between $13$ and $18$. The players did not know the total number of rounds they were going to play. For each round, subjects were randomly assigned different opponents, and nobody knew who they were playing against. In each round of the game, the players had 40 seconds to make their action choice. If they did not choose anything, a random choice was generated by the system (and saved in our database, properly labeled to be discarded in the analysis). After a player had made a decision in a particular round, she had to wait until all other players were done too, before obtaining the outcome of the round and proceeding to the next game round (Fig.~\ref{fig:tutorial_1} j). Finally, in order to encourage the experimental subjects' decisions with real material (economic) consequences, they were informed that they would receive lottery tickets proportionally to the payoff they accumulated during the rounds of dyadic games they played. The four prizes in the corresponding lottery were coupons redeemable at participating neighboring stores, worth 50 euros each.

\section*{S.3.\ Translated transcript of the tutorial and feedback screen after each round}
\label{sec:instructions}

Before the experiment started, and for each group of subjects, we showed them a tutorial in the same tablets used to play the game. The format and presentation of the game examples used in the tutorial were identical to those of the real experiment.
We present next the translation into English of the text from every screen of the tutorial (the original was made available to the participants in Castilian/Spanish and Catalan.

\textbf{Tutorial Screen \#1. See Fig.~\ref{fig:tutorial_1}(a).} \emph{Welcome to Dr. Brain. The game, designed to study how we make decisions, is made of several rounds with different opponents located in the DAU.
During the experiment we don't expect you to behave in any particular way: there are no wrong nor incorrect answers. You will simply have a limited time to make your decisions.
In these next screens we will teach you how to play Dr. Brain. Use the side arrow keys to move within the tutorial, and when you are done you will be able to start the rounds.
This game has been thought by scientists from the Universitat of Barcelona (UB), Universitat Rovira i Virgili (URV), Instituto de Biocomputaci\'on and Sistemas Complejos (BIFI)-Universidad de Zaragoza (UZ) and Universidad Carlos III in Madrid (UC3M). It is an experiment to study and understand how we humans make decisions.}

\textbf{Tutorial Screen \#2. See Fig.~\ref{fig:tutorial_1}(b).} \emph{The rules of Dr. Brain. It is important that you don't talk to other players during the experiment. Keep focused!
The decisions made during the experiment and the accumulated points will determine your chances of wining prizes: the more points, the more tickets you will get for the raffle.
If you leave the game while it is in progress, you won't be able to come back in!}

\textbf{Tutorial Screen \#3. See Fig.~\ref{fig:tutorial_1}(c).} \emph{This is the screen you will see when the rounds of the game start. In each one of them, we will assign you a random partner to play.}

\textbf{Tutorial Screen \#4. See Fig.~\ref{fig:tutorial_1}(d)}. \emph{Each round has a table that represents your opponent's possible actions as well as yours. Your opponent and you will follow the same rules in the round. In this way, depending on what each one of you choose, you will win more or less.
The rows represent your choice, the columns represent your opponent's. For each choice, it is listed how much you will win, and how much your opponent will.}

\textbf{Tutorial Screen \#5. See Fig.~\ref{fig:tutorial_1}(e).} \emph{Pay attention, the tables may change from round to round, and the rules may be different. You may win more or less points, o what seemed more interesting may be different now.}

\textbf{Tutorial Screen \#6. See Fig.~\ref{fig:tutorial_1}(f).}
\emph{To play you must choose one of the two options, represented by a color. Your opponent plays following the same rules as you, described in the table, but you won't know his choice until after the end of the round.}

\textbf{Tutorial Screen \#7. See Fig.~\ref{fig:tutorial_1}(g).}
\emph{Every round of the game lasts 40 seconds, you have to choose one of the two actions during that time. If you don't choose anything, the computer will do it for you randomly and you will move on to play the next round. Don't worry, 40 seconds is plenty of time!}

\textbf{Tutorial Screen \#8. See Fig.~\ref{fig:tutorial_1}(h).}
\emph{Example: If you pick RED and your opponent picks GREEN. You (red) win 8 and your opponent (green) wins 6.}

\textbf{Tutorial Screen \#9. See Fig.~\ref{fig:tutorial_1}(i).}
\emph{Example: If you pick PURPLE and your opponent picks YELLOW. You (purple) win 11 and your opponent (yellow) wins 0. If your adversary chooses... If you choose... You win... He wins... What do you choose?}

\textbf{Feedback Screen after a typical round of the game. See Fig.~\ref{fig:tutorial_1}(j).}
\emph{Almost there, thanks for your patience! You and your opponent have both chosen YELLOW. You and your opponent have earn 5 each. Next game starts in... (countdown)}

\section*{S.4.\ Other experimental results}

\subsection*{S.4.1.\ Fraction of cooperation by age and gender}
\label{subsec:demographics_coop_age_gender}

We did not find any significant differences in the fraction of cooperative actions in the whole $(T,S)$-plane by age when separating young players ($\leq 15$ years old) from adults ($>16$ years old) (see Fig.~\ref{fig:heatmaps_young_old}) nor between males and females (see Fig.~\ref{fig:heatmaps_male_female}).

\subsection*{S.4.2.\ Fraction of cooperation by game round}
\label{sec:cooperation_by_round}

We did not observe large differences in the fraction of cooperative actions in the whole $(T,S)$-plane when separating by game round, with the exception of the first few rounds of the session (see Fig.~\ref{fig:heatmaps_by_round} for heatmaps of cooperation and Fig.~\ref{fig:heatmaps_by_round_relative_diff} of heatmaps of relative differences in cooperation).

\subsection*{S.4.3.\ Number of actions per $(T,S)$-plane point and Standard Error of the mean fraction of cooperation}
\label{sec:sem_coop}

The total number of actions generated by our $541$ subjects was $8,366$. The $(T,S)$-plane was discretized into a $11 \times 11$ lattice, and the $(T,S)$ point for any given pair of opponents and for any given round was randomly generated in such a way that subjects had uniform probability to be assigned to any point in the $(T,S)$-plane. Thus, the average number of actions per $(T,S)$ point is $69$. In Fig.~\ref{fig:Num_actions_cooperation_map} we show the total number of actions per point in the $(T,S)$-plane for all subjects. 

On the other hand, in Fig.~\ref{fig:SEM_cooperation_map}, we show the Standard Error of the mean fraction of cooperative actions for all the actions and all the players in the experiment, for the whole (T,S)-plane. We observe that the values for the Standard Error of the mean are uniformly distributed across the entire (T,S)-plane, except for the upper-left triangle of the HG, where the error is clearly lower than in the rest of the regions. This seems to indicate that at a population level, most people chose the same action at least in that particular region.

\subsection*{S.4.4.\ Time evolution of the fraction of cooperation}
\label{sec:time_evol_coop}

Our experiment was designed to avoid learning or memory effects as much as possible, making each subject play knowingly in different game conditions and against different anonymous opponents in every round. In the left panel of Fig.~\ref{fig:Coop_vs_time}, we show the average fraction of cooperative actions as a function of the round number over the whole population, and we observe how there is only a very small decline in cooperation as the round number increases, specially during the first two or three rounds. Also, note that the dispersion of the values is larger in the last few rounds, since every subject play a random total number of rounds between 13 and 18 rounds. Similarly, we show in the right panel of Fig.~\ref{fig:Coop_vs_time} the average fraction of cooperative actions as a function of the round number, separating the actions into the different games. In this case we do observe a small decline of cooperation in the case of the Prisoner's Dilemma (PD) and the otherft (SG), and a small increase in cooperation in the Harmony (HG), while the fraction of cooperative actions doesn't show any particular trend for the Stag Hunt (SH).

\subsection*{S.4.5.\ Rationality and Risk aversion}
\label{sec:rationality}
We measure the level of rationality (only under the assumption of self-interest) among our subjects using only their actions in the Harmony and/or Prisoner's Dilemma games. According to Game Theory, the rational action in the Harmony game is to cooperate, while in the Prisoner's Dilemma it is to defect.

In Fig.~\ref{fig:hist_rationality} we show the distributions of the fraction of rational actions chosen by the subjects in the Harmony game (HG), in the Prisoner's Dilemma (PD), and in both games combined, along with the corresponding mean values among the population (vertical purple lines). We observe that an important subset of individuals presents a fraction of rational actions near $1.0$ (around $50\%$ of subjects when calculated with either game independently, and around $30\%$ when calculated with both games combined).
However, there are also some others that act irrationally (around $5\%$ or $10\%$ as calculated with either game). Note that the average value of rationality of the whole population when both games are considered in the statistics, is around $75 \%$ (see purple vertical lines in Fig.~\ref{fig:hist_rationality}).

Moreover, we checked the time evolution of the fraction of rational actions in the population, as defined by their actions in the Harmony (HG) and Prisoner's Dilemma (PD) games together, and independently (Fig.~\ref{fig:time_evol_rationality}), and we do not observe any significant increase or decrease of rationality as a function of the round number in any case.

Regarding the definition of risk-aversion, we choose to define it as the number of cooperative actions in the SG together with the number of defective actions in the SH (over the total sum of actions in both quadrants for a given player). The rationale behind such a combined measure of risk aversion is the avoidance of the bias of pure cooperativeness: were we to measure risk aversion only in the SH (instead of combining both SH and SG), for a group that defects a lot everywhere in the $(T,S)$-plane, it would appear as if they are more risk averse than they really are, while a mostly cooperative group would appear as less risk averse than they really are. A similar reasoning would apply to only using the SG quadrant for the measure, and therefore we  have looked at the actions in both the coordination and anti-coordination games together.

In Figure \ref{fig:risk_aversion} we represent the average values of risk-aversion according to this definition, for each one of the phenotypes, and the population as a whole. While  Envious, Trustful, and Un players exhibit intermediate levels of risk aversion ($0.52$, $0.52$ and $0.54$, respectively),  Pessimists exhibit a significantly higher value ($0.73$), consistent with their fear of facing the worst possible outcome and their choice of the best worst-case scenario. In contrast, the Optimist phenotype shows a very low risk aversion ($0.32$), in agreement with the fact that they aim to obtaining the maximum possible payoff, risking the possibility that their counterpart do not work with them towards that goal.

\subsection*{S.4.6.\ Response times}
\label{sec:response_times}

We have also examined the response times of the individuals in our experiment, separating the data by cooperation/defection actions, and as a function of the round number. Fig.~\ref{fig:response_time} shows that the average response time is around 15 seconds. We did not find any dependence with the round number nor with the type of action. Finally, Fig.~\ref{fig:response_time_hist} displays the distributions of response times for all individuals, for each of the two possible actions.

\subsection*{S.4.7.\ Clustering Analysis}
\label{sec:clustering_analysis}

We hypothesized that there are distinct, well-defined types of individuals (or phenotypes) in our dataset, that can be told apart by using an unsupervised clustering algorithm.
Hence, we run a $K$-means clustering algorithm on our data (using the Scikit-learn Python package) to analyze its clustering structure. We represent each participant in the dataset by a four-dimensional vector, corresponding to her average fraction of cooperative actions in each one of the four dyadic games (Prisoner's Dilemma, Stag Hunt, Snowdrift and Harmony). 

The $K$-means unsupervised clustering algorithm groups the data into a user-defined number of clusters, by both minimizing the dispersion within each cluster and maximizing the distance between the centroids of each pair of clusters. For a given number of clusters, $k=2,3,4,...,20$, we run the algorithm $200$ times on our data (with different seeds for the algorithm in every run), and obtain the average value of the BD-index (see subsection below for formal definition), which is a measure of how optimal is that $K$-scheme. This way we can pick which one is the best cluster scheme. In Fig.~\ref{fig:robustness} we show the average value and the Standard Deviation (SD) of the DB-index, as a function of the number of clusters in the partition. This representation will have a minimum around the optimum number of clusters for a given dataset. Conversely, it would be monotonically decreasing if the data set lacks any significant cluster structure.

We found that there is an optimum around a scheme with 5 or 6 clusters (black line in Fig.~\ref{fig:robustness}). However, due to the fact that the SD is considerably smaller for 5 than for 6 (which indicates that the partition schemes found in different realizations of the algorithm for $k=5$ are much more similar to each other in terms of their corresponding DB-index, than in the case of $k=6$), we pick $k=5$ as our optimum clustering partition. Note also that the SD is very large for any partition with 6 or more clusters, which also points to the lack of robustness of those partition schemes.

It is also important to mention that this clustering approach does not allow us to compare our results against the 'ground truth', since that is unknown to us. We can only test for its robustness, and we do this in multiple ways. We present the results from the same algorithm, also run $200$ times, but this time on a randomized version of our data. This data randomization is done as follows: we take the $8,366$ actions of the $541$ subjects and create an 'action pool' with them. From this pool of data we draw (with replacement) to obtain the new, randomized sets of actions for each person, in such a way that we preserve the number of times each subject has played and the particular $(T,S)$ points she played in, but now her actions are randomized. With this randomization procedure we preserve the average fraction of cooperative actions in the population, but destroy any possible correlations among the actions of any given subject. 
Note in Fig.~\ref{fig:robustness} that with the randomized version of the data (green line), there is no local minimum for the DB index, and the best partition would be to have as many clusters as possible, which is an indication of the lack of internal structure of the randomized data.

On the other hand, and recalling that the cooperation patterns in the heatmap for all users seems to be a little less clear during the first few rounds (while the subjects seem to be picking up the mechanics of the experiment), than during the rest of the experiment (see Fig.~\ref{fig:heatmaps_by_round}), we also test the clustering structure of our data when removing the first couple of rounds for every subject. In this case, we observe that the cluster structure is even clearer, with an even more significant minimum at $k=5$ clusters, as indicated by the DB-index (Fig.~\ref{fig:robustness}, red line).

On the other hand, we also wanted to test the robustness of our clustering analysis against data perturbations, specifically by running it on just a subset of the original data. In order to do so, we run the algorithm $200$ times again, but in each realization we exclude a given number of players and all their actions, randomly chosen (that is to say, we perform a leave-p-out analysis, for different values of $p$). We do this for a scheme with $k=2,3,4,...,20$ clusters, and leaving out $p=100, 300, 400$, and $450$ subjects (out of the total $541$), and calculate again the average DB index for them.
In Fig.~\ref{fig:DBindex_leave_p_out} we show the results from the leave-p-out procedure as they compare to the original data (the black dashed line in Fig.~\ref{fig:DBindex_leave_p_out}). We observe that the results of the $K$-means analysis in our data are very robust when randomly removing $p \leq 300$ subjects from the original set and all their actions (that is up to 55\% of the data): we observe that the optimum in the DB index remains around the same value $k=5$. However, the SD is larger for all the leave-p-out cases, and for any given $k$ or $p$, than for the analysis perform over the original data set. This variability gets larger the more data is randomly excluded. 
Of course, if too much of the data is removed ($p \geq 300$ subjects), the $K$-means algorithm is no longer able to retrieve the original optimum cluster structure, as can be inferred from the gradual disappearance of the local minimum in Fig.~\ref{fig:DBindex_leave_p_out} as $p$ increases. We remind the reader that a data set lacking any cluster structure would render a monotonically decreasing DB index as a function of the number of clusters.

\subsection*{S.4.8.\ DB index}
\label{subsec:DBindex}

The Davies-Bouldin index, or DB index (30), is a metric for evaluating and comparing clustering algorithms. It is minimized by the optimum clustering scheme, that is to say, by the partition in a number of clusters such that it presents the minimum dispersion within each cluster, and the maximum distance between all pairs of clusters.
In particular, this metric performs an internal evaluation, that is, the validation of the goodness of the clustering partition is made using quantities inherent to the data set. Hence, it does not do a validation against the 'ground truth'. We picked this particular validation method because in this context there isn't a known ground truth for types of players (or 'phenotypes').

Given a certain scheme or partition in $N$ clusters, let $C_i$ be a cluster of vectors, and let $\vec{X_\ell}$ be an n-dimensional feature vector that represents subject $\ell$ (in our particular case, $n=4$ dimensions), who is assigned to cluster $C_i$. The dispersion $S_i$ within cluster $C_i$ is calculated as: 
\begin{equation}
 S_{i} = \frac{1}{T_{i}}\sum_{\ell=1}^{T_{i}} \lVert \vec{X_{\ell}}-\vec{A_{i}}\rVert,
 \label{eq:dispersion}
\end{equation}
where $\vec{A_i}$ is the centroid of cluster $C_i$, $\lVert \vec{X_{\ell}}-\vec{A_{i}}\rVert$ denotes the Euclidean distance between the vector $\vec{X_\ell}$ and the centroid $\vec{A_i}$, and $T_i$ is the size of cluster $C_i$ (that is, the number of subjects assigned to that cluster).

Then for each pair of clusters $i$ and $j$, we define the matrix
\begin{equation}
 R_{ij} = \frac{S_i+S_j}{M_{ij}},
 \label{eq:separation_matrix}
\end{equation}
where $M_{ij}=\lVert \vec{A_{i}}-\vec{A_{j}}\rVert$ is the separation between clusters $i$ and $j$ (that is, the distance between their corresponding centroids).

Thus, we can define the DB index as
\begin{equation}
 DB = \frac{1}{N}\sum_{i=1}^{N}D_i,
 \label{eq:level_of_cooperators}
\end{equation}
where $D_i=\max\limits_{i \neq j} R_{ij}$.

\subsection*{S.4.9.\ Normalized Mutual Information Score}
\label{subsec:mutual_info}

In order to compare the consistency between two independent runs of the $K$-means algorithm in terms of the individuals' composition of the clusters obtained, we use the Normalized Mutual Information Score (37), as implemented in the Python package SciKit Learn).

The Mutual Information is a measure of the similarity between two clustering (or labeling) systems $U$ and $V$ of the same data into disjoint subsets, and it is given by the relative entropy between the joint distribution and the product distribution. Mutual Information between clustering systems $U$ and $V$ is then defined as:

\begin{equation}
MI (UV) = \sum_{i=1}^U \sum _{j=1}^V P(i,j) log \frac{P(i,j)}{P(i)P'(j)} .
 \label{eq:mutual_info}
\end{equation}

Where $P(i)$ is the probability of a random sample occurring in cluster $U_i$ and $P'(j)$ is the probability of a random sample occurring in cluster $V_j$.

To obtain a Normalized Mutual Information Score in such a way that it is bounded between 0 (no mutual information) and 1 (perfect correlation), the Mutual Information is normalized by $\sqrt{H(U) * H(V)}$, being $H(U)$ the entropy of the  clustering system $U$, and $H(V)$ that of clustering $V$.

Note that this metric is independent of the absolute values of the labels: a permutation of the class or cluster label values will no��t change the score value in any way, and furthermore, it is symmetric, since switching the labels from clustering system $U$ to clustering system $V$ will return the same score value. 

In Fig.~\ref{fig:mutual_info} we present the average value of Normalized Mutual Information, $\left< MI \right>$, for any number of clusters in the original data, and in the randomized version of the data, over $2,000$ runs of the algorithm. We proceed as follows: we perform ($1,000$) pair-wise comparison of clustering schemes obtain in different runs, calculating its corresponding score in each case, so then we can obtain an average.
We observe how the score is significantly higher in the case of the actual data, than when comparing with the results from a randomized version of the data (for example, $\left< MI \right>=0.97$, $SD:0.03$, vs $\left< MI \right>=0.59$, $SD:0.18$ for actual and randomized results at $k=5$ clusters), which indicates that the individuals composition of the clusters in any two runs of the algorithm on the real data are extremely correlated, but it is not the case for two runs over randomized data. Finally, we also report that the score is at its highest value for $k=5$ clusters.

\subsection*{S.4.10.\ Age and gender by phenotype}
\label{subsec:age_gender_cluster}

The average (SD) age by phenotype is: for the Envious is $29.9 (13.9)$; $32.5 (13.7)$ for the Optimist; $32.0 (16.8)$ for the Undefined; $32.29 (14.1)$ for the Cooperators, and $30.7 (13.8)$ for the Pessimist. We do not observe significant differences on the average age among different phenotypes nor with respect with the population average ($31.3$, SD: $14.3$). 

In Fig.~\ref{fig:age_distrib_clusters} we present the age distributions by phenotype, as they compare to the distribution for the whole population. We do not observe significant differences for any of the distributions by phenotype when comparing with that for the whole population nor by doing pair-wise comparisons of different phenotypes. The corresponding p-values for the KS-test (used to compare the probability distribution of two samples) of each possible pairwise combinations are non-significant: Envious vs Optimist: 0.31; Envious vs Undefined: 0.29; Envious vs Trustful: 0.32; Envious vs Pessimist: 0.81; Optimist vs Undefined: 0.57; Optimist vs Trustful: 0.99; Optimist vs Pessimist: 0.64; Undefined vs Trustful: 0.71; Undefined vs Trustful: 0.71; Undefined vs Pessimist:0.68; Trustful vs Pessimist: 0.67. 
Similarly, the p-values for all comparisons between clusters and the whole population are non-significant: Optimist vs all: 0.88; Envious vs all: 0.79; Undefined vs all: 0.68; Trustful vs all: 0.88; Pessimist vs all: 0.81.

The percentage of males for each phenotype is: $67\%$ among the Envious, $64\%$ among the Optimist, $64\%$ among the Undefined, $61\%$ among the Cooperators and $64\%$ among the Pessimists (while the percentage of males for the whole populations is $64\%$). The z-scores of the comparison of gender distributions of each cluster vs the whole population by bootstrapping are all non-significant: Envious vs all: -0.036; Optimist vs all: -0.460; Undefined vs all: -0.260 ; Trustful vs all: -0.646; Pessimist vs all: -0.132.

\subsection*{S.4.11.\ Differences between experimental and numerical behavioral heatmaps}
\label{subsec:differences_heatmap_coop}

Assuming that each subject in our study plays using one and only one of the behavioural rules or phenotypes, and preserving the relative fractions of each one of them present in the population as found by the clustering algorithm, we can compute the differences between experimental and numerical (or inferred) behavioral heatmaps for each phenotype. In 
Fig.~\ref{fig:differences_experiment_numerical} it can be seen that, even if occasionally the difference can reach up to 4 SD units for a particular $(T,S)$ point, there is no systematic bias in any of the different heatmaps. The average difference in the aggregate case is of $1.39$ SD units, while the difference by phenotype are:  $1.91$ SD units for Envious, $1.85$ SD units for Optimist, $2.14$ SD units for Pessimist, $1.79$ SD units for Trustful, $1.12$ SD units for Undefined. Thus none of the phenotypes presents an average difference beyond the 99\% Confidence Interval ($2.575$ SD units). Indeed, only Pessimists present an average difference out of 95\% Confidence Interval ($1.96$ SD units), the rest are below such standard threshold. We thus clearly show that the aggregation of the behavior of our volunteers into the proposed phenotypes is not significantly different from what we have obtained in the experiment.

\subsection*{S.4.12.\ Dependence of cooperation on $S-T$}

Inspired by the population-level observation about the patterns described by lines parallel to $S=T$, and the fact that the population as a whole does not seem to distinguish between SH and SG, we studied cooperation as a function of the combined variable $(S-T)$. The results, represented in Fig.~\ref{fig:cooperation_vs_SminusT_all}, show a remarkable collapse of all curves into a single one, indicating that the aggregate cooperation level can be described by $(S-T)$, as previously pointed out by Rapaport (11, 13). In this respect, it is worth noting that $(S-T)$ represents the maximum possible payoff difference for any game. For very negative values of $(S-T)$, which corresponds to the PD game, the levels of cooperation are low but not zero, while for positive values (corresponding to HG) they are high, with intermediate, increasing values of cooperation for the region $(S-T) \in [-10,0]$, which roughly corresponds to a combination of the coordination and the anti-coordination games. This suggests that competition, in the sense of ending up being better off than one's counterpart, may be important for our experimental subjects.

Further, we check whether these results are reproduced from our interpretation of the clustering results and the corresponding simulations. In Fig.~\ref{fig:S-T_final} we plot together the results obtained from numerical simulations that use the experimentally obtained classification. As shown, by simply using the right fraction of each phenotype (behavioral rules) in the population, we can recover the observed diagonal symmetry, thus further confirming our $5$-phenotype hypothesis.

\bibliographystyle{Science}

\clearpage

\begin{figure}[h]  
\begin{center}$
\begin{array}{cc}
\includegraphics[width=0.70\textwidth]{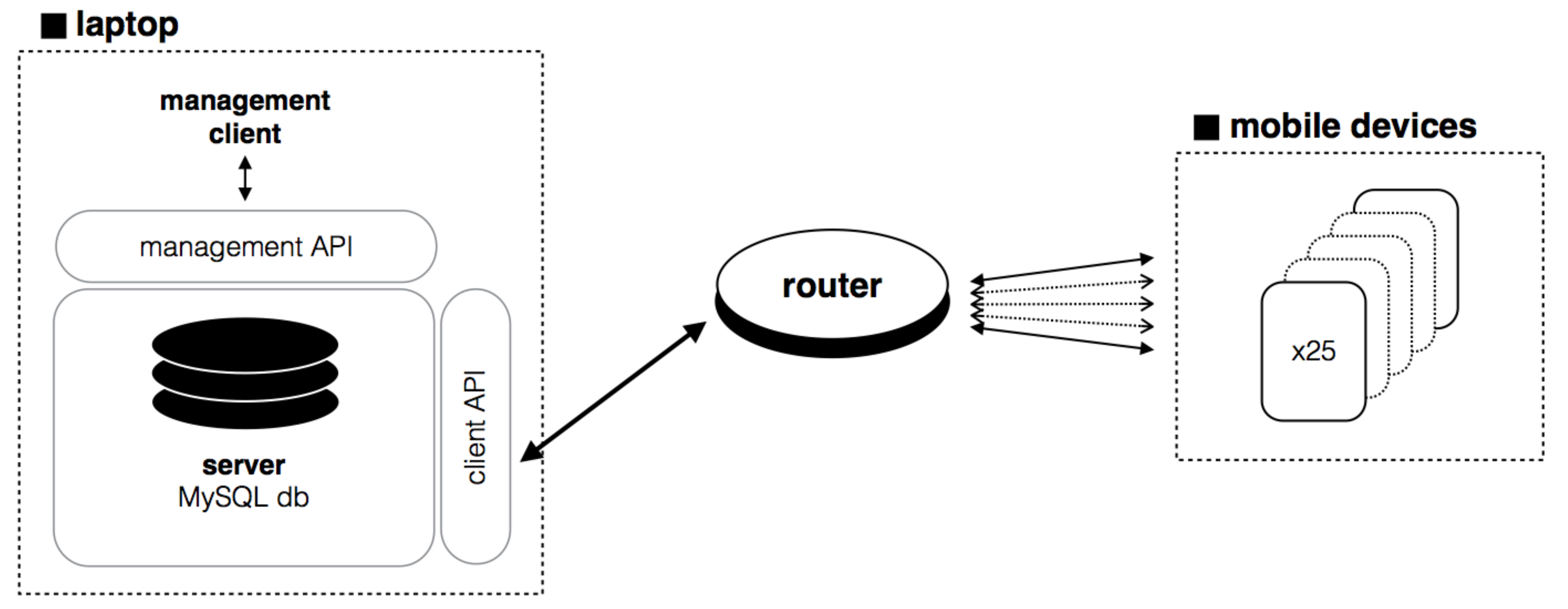} 
\end{array}$
\end{center}
\caption{System architecture.}\label{fig:system_architecture}
\end{figure}

\begin{figure}[h]   
\begin{center}$
\begin{array}{cc}
\includegraphics[width=0.75\textwidth]{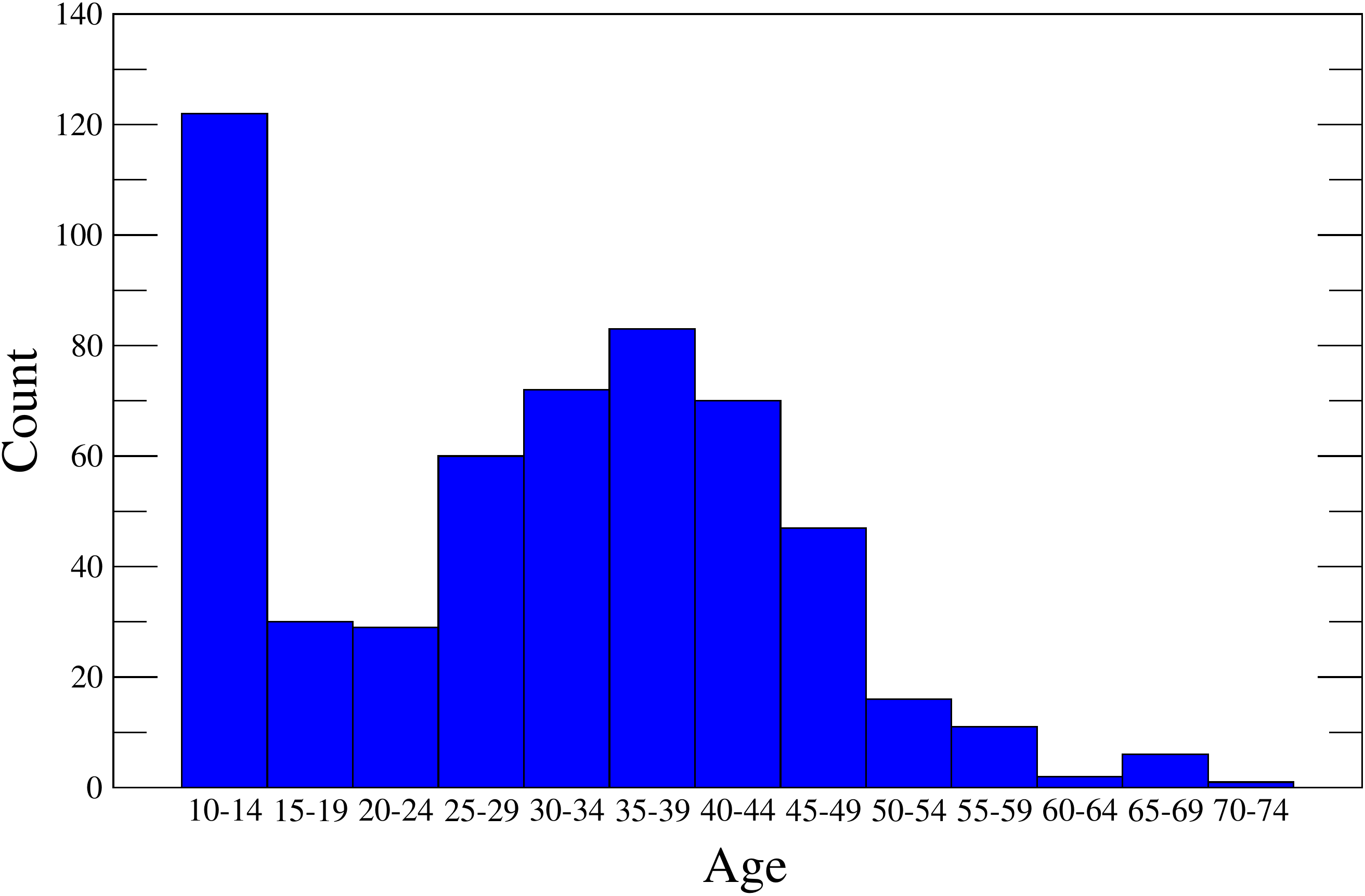} 
\end{array}$
\end{center}
\caption{Age distribution of the participants in our experiment.}\label{fig:age_distrib}
\end{figure}

\clearpage

\begin{figure*}[h]  
\begin{center}$
\begin{array}{lll}
(a) \includegraphics[width=0.3\textwidth]{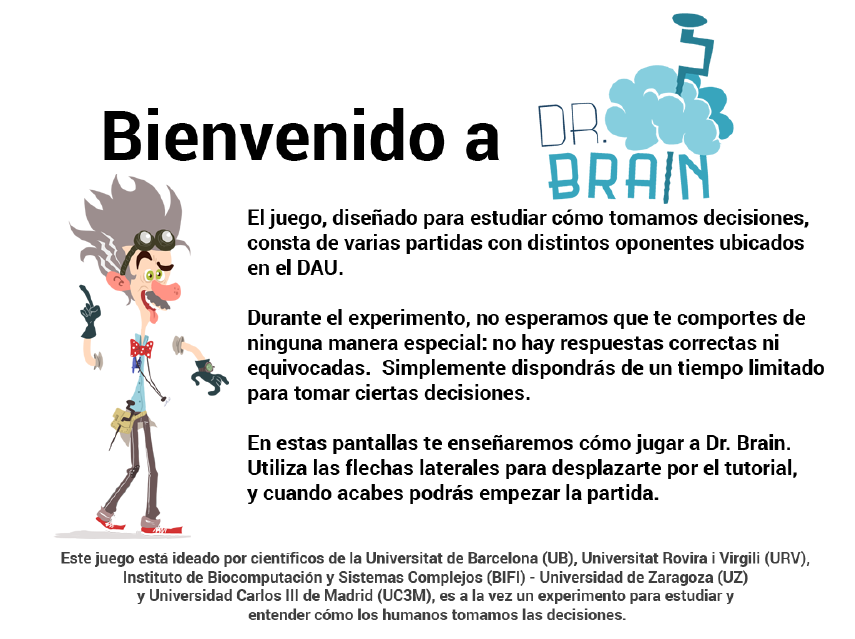}

\vspace{2mm} 

(b)
\includegraphics[width=0.3\textwidth]{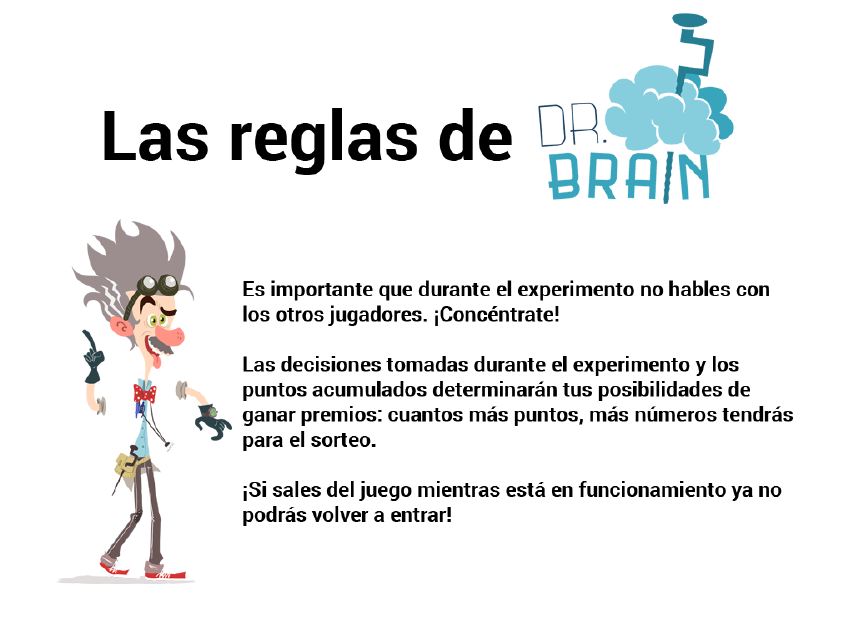}

\vspace{2mm}

(c) 
\includegraphics[width=0.3\textwidth]{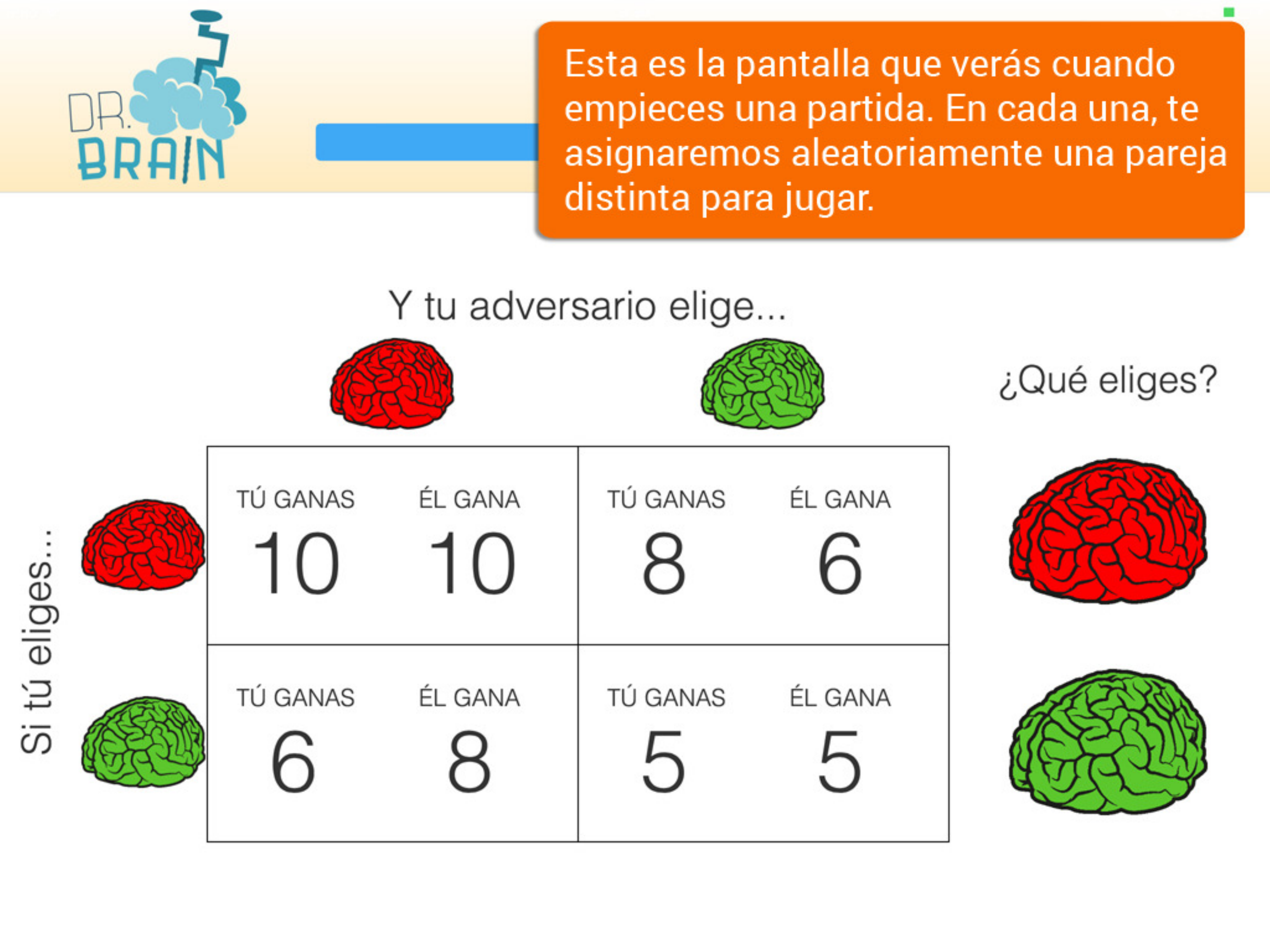}\\

\vspace{2mm} 

(d)
\includegraphics[width=0.3\textwidth]{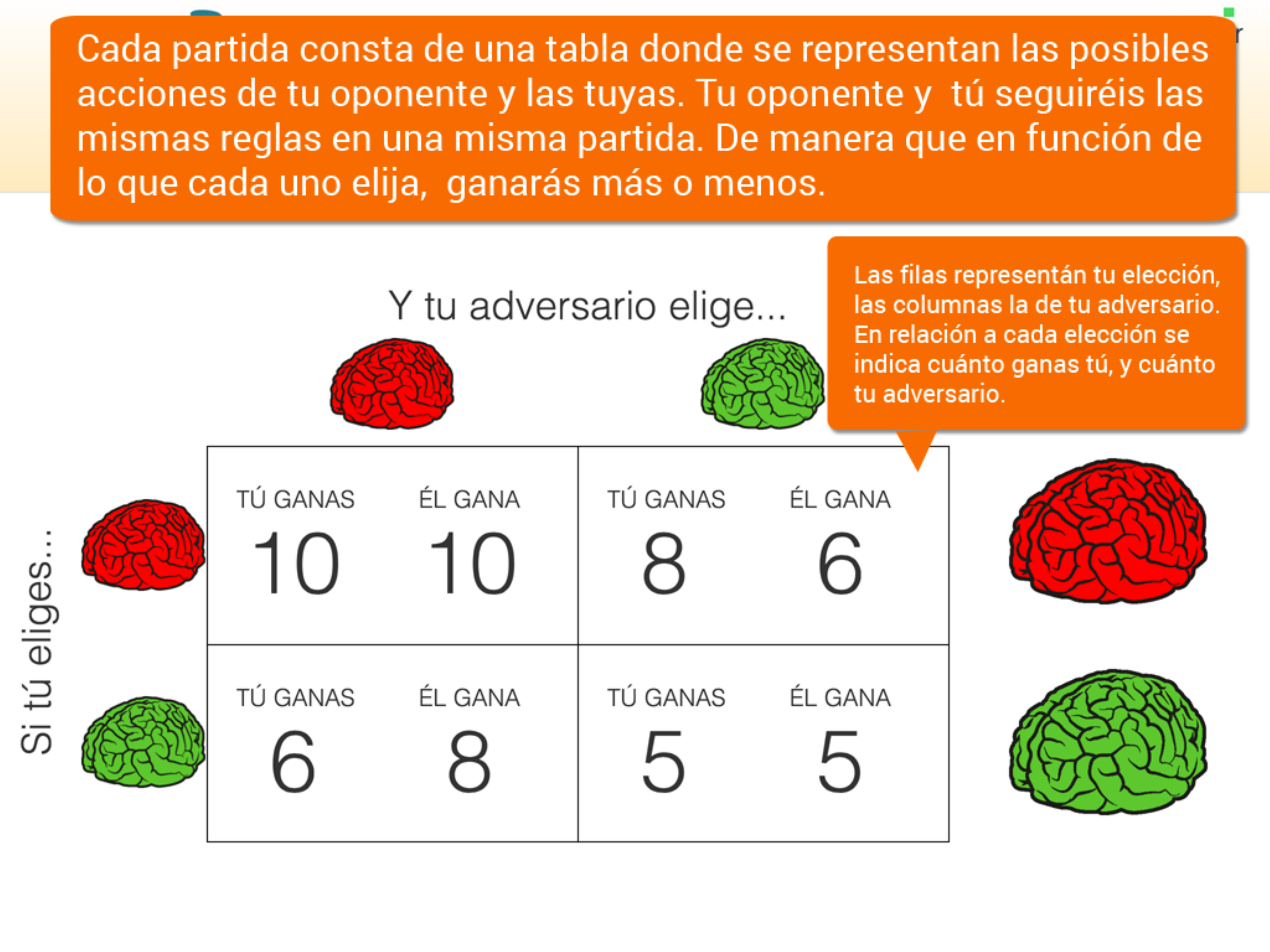}

\vspace{2mm} 

(e)
\includegraphics[width=0.3\textwidth]{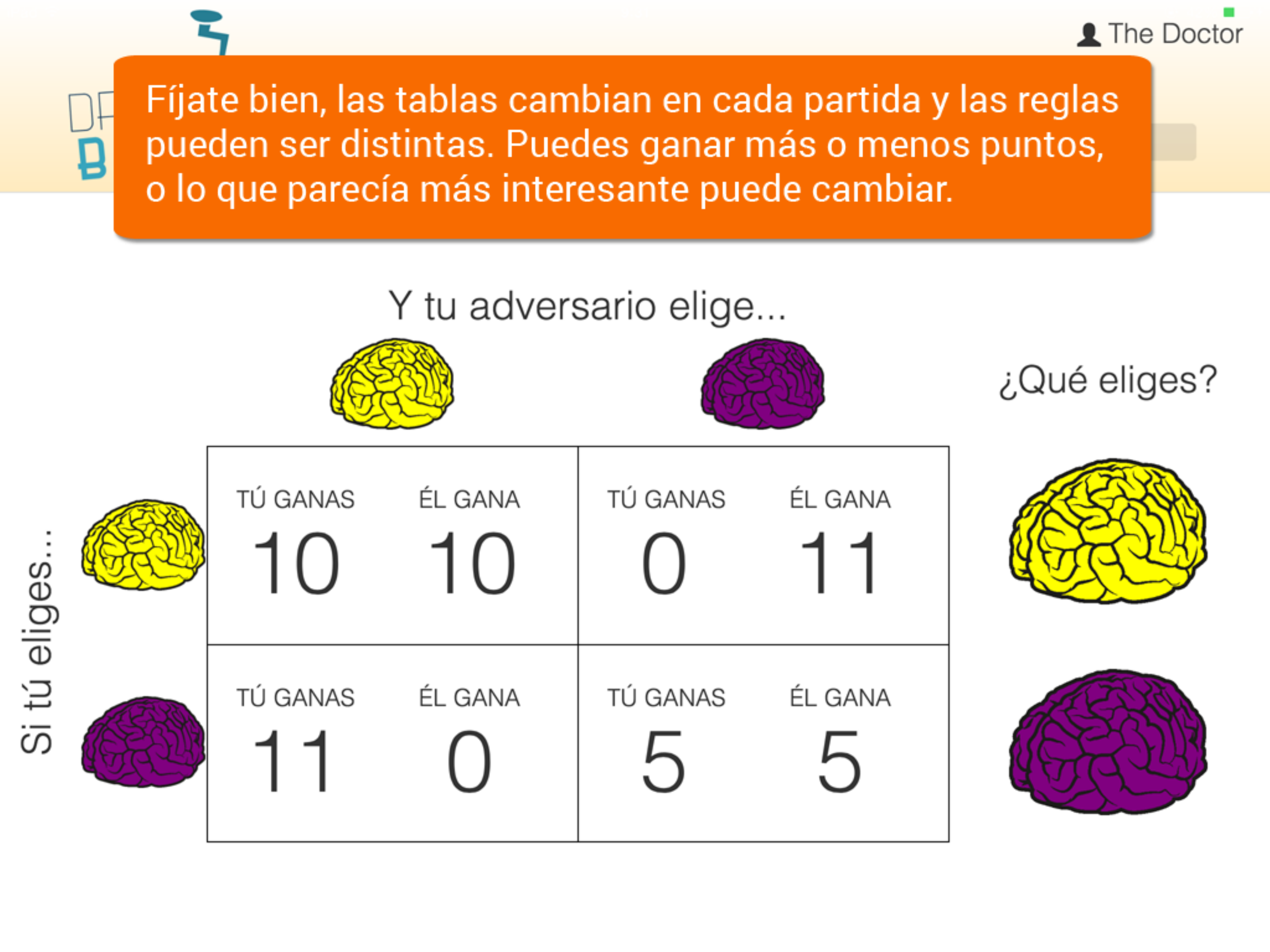} 

\vspace{2mm} 

(f)
\includegraphics[width=0.3\textwidth]{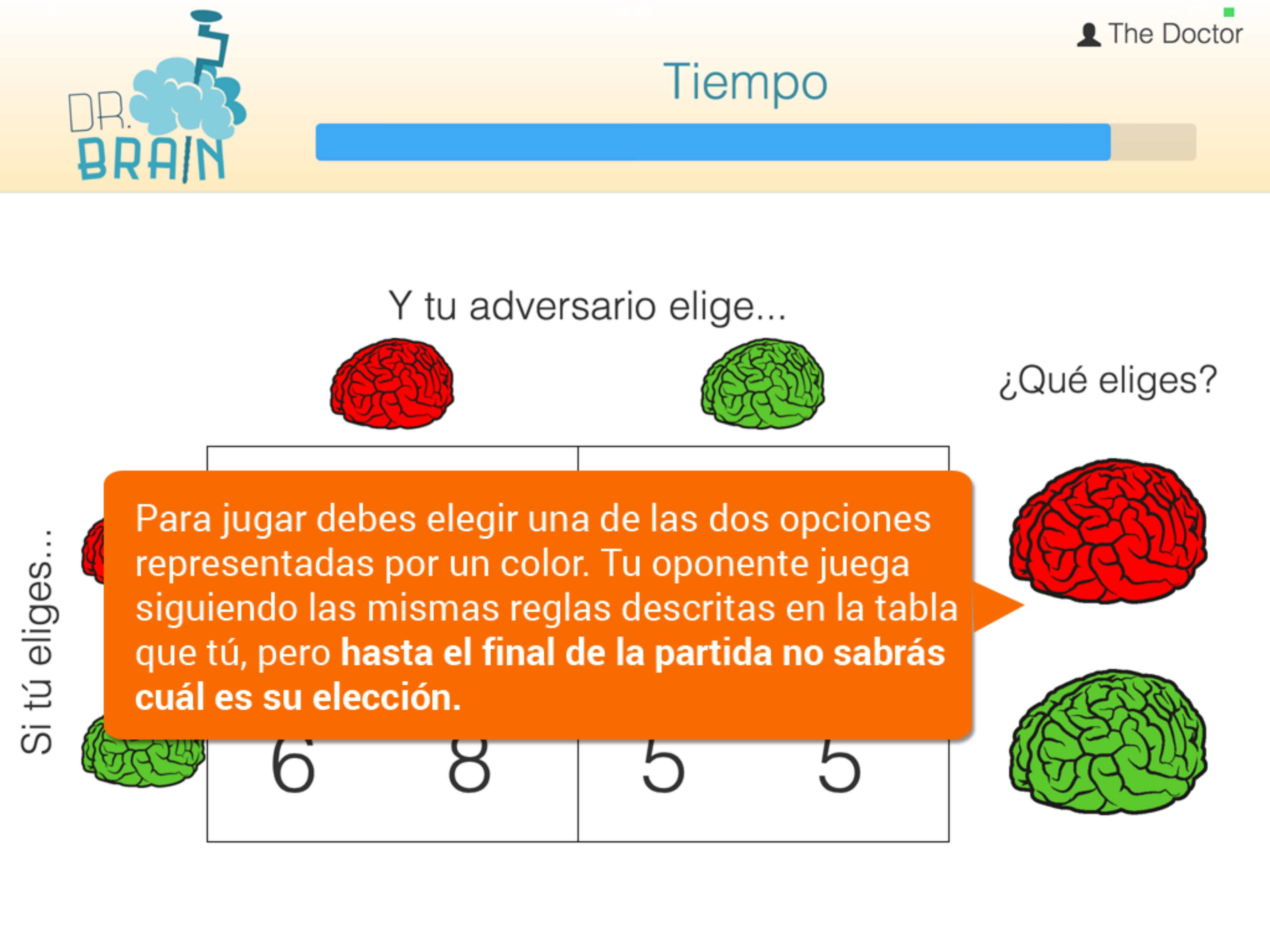}\\

\vspace{2mm} 

(g)
\includegraphics[width=0.3\textwidth]{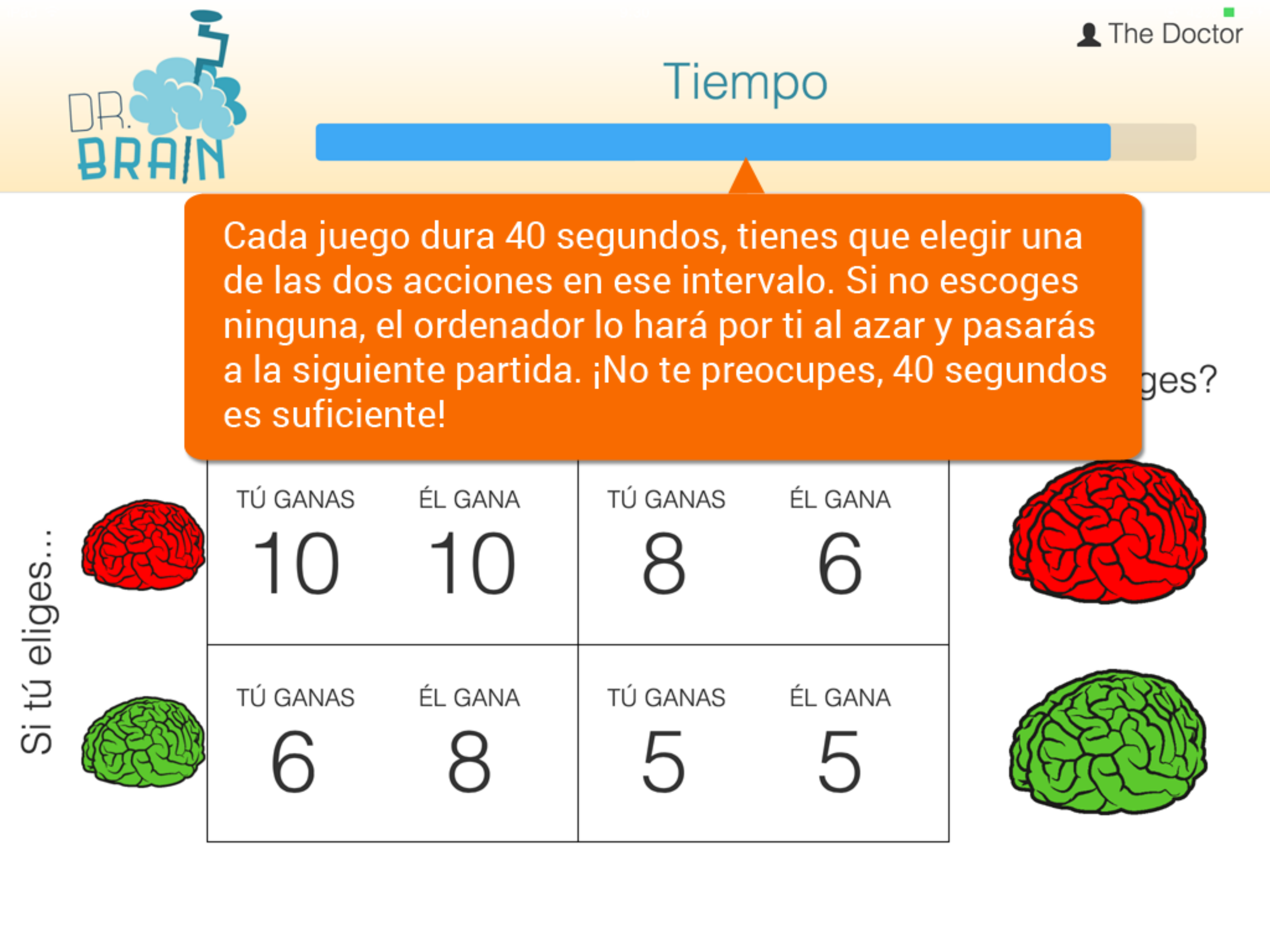} 

\vspace{2mm} 

(h)
\includegraphics[width=0.3\textwidth]{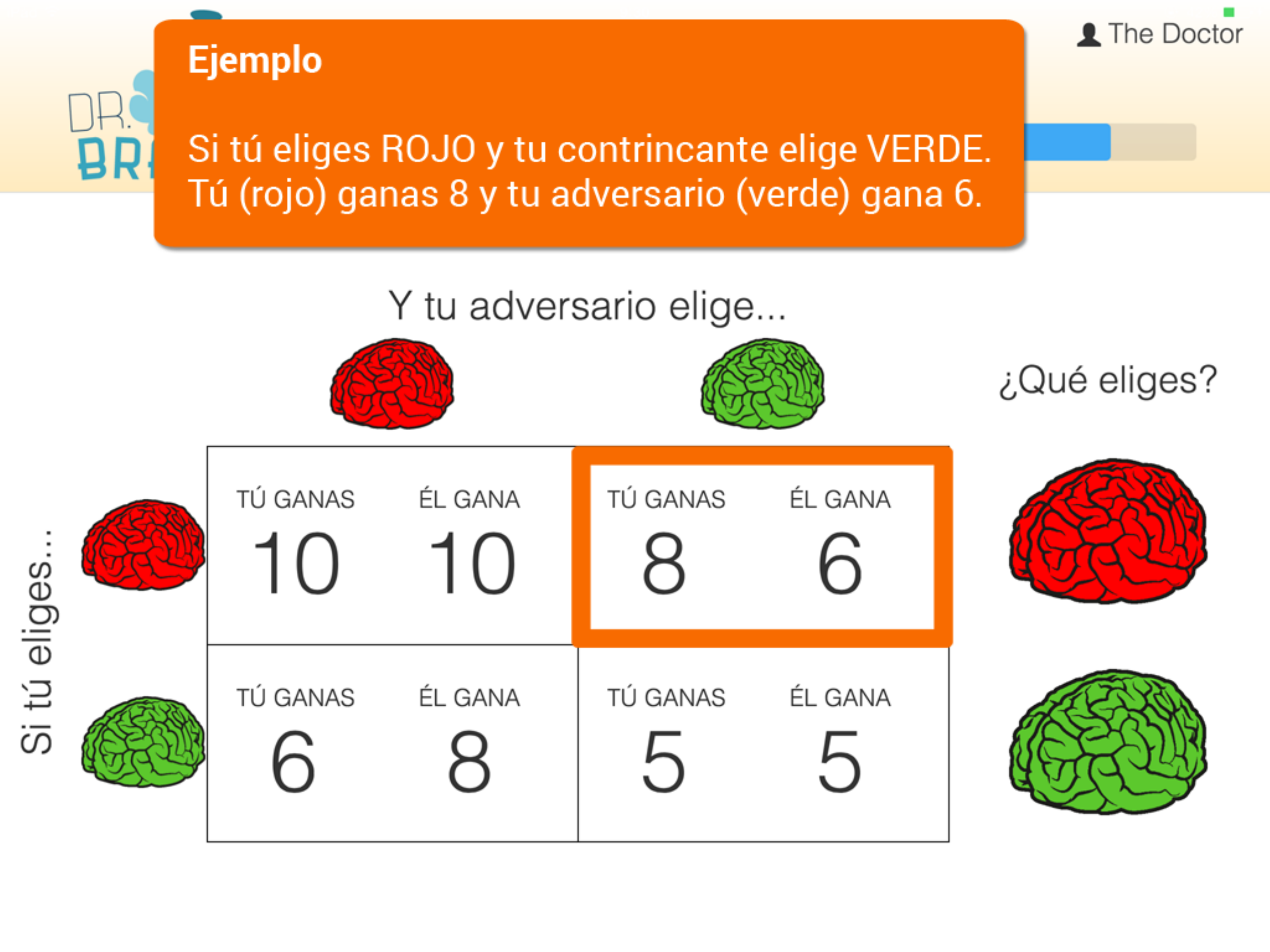}

\vspace{2mm} 

(i)
\includegraphics[width=0.3\textwidth]{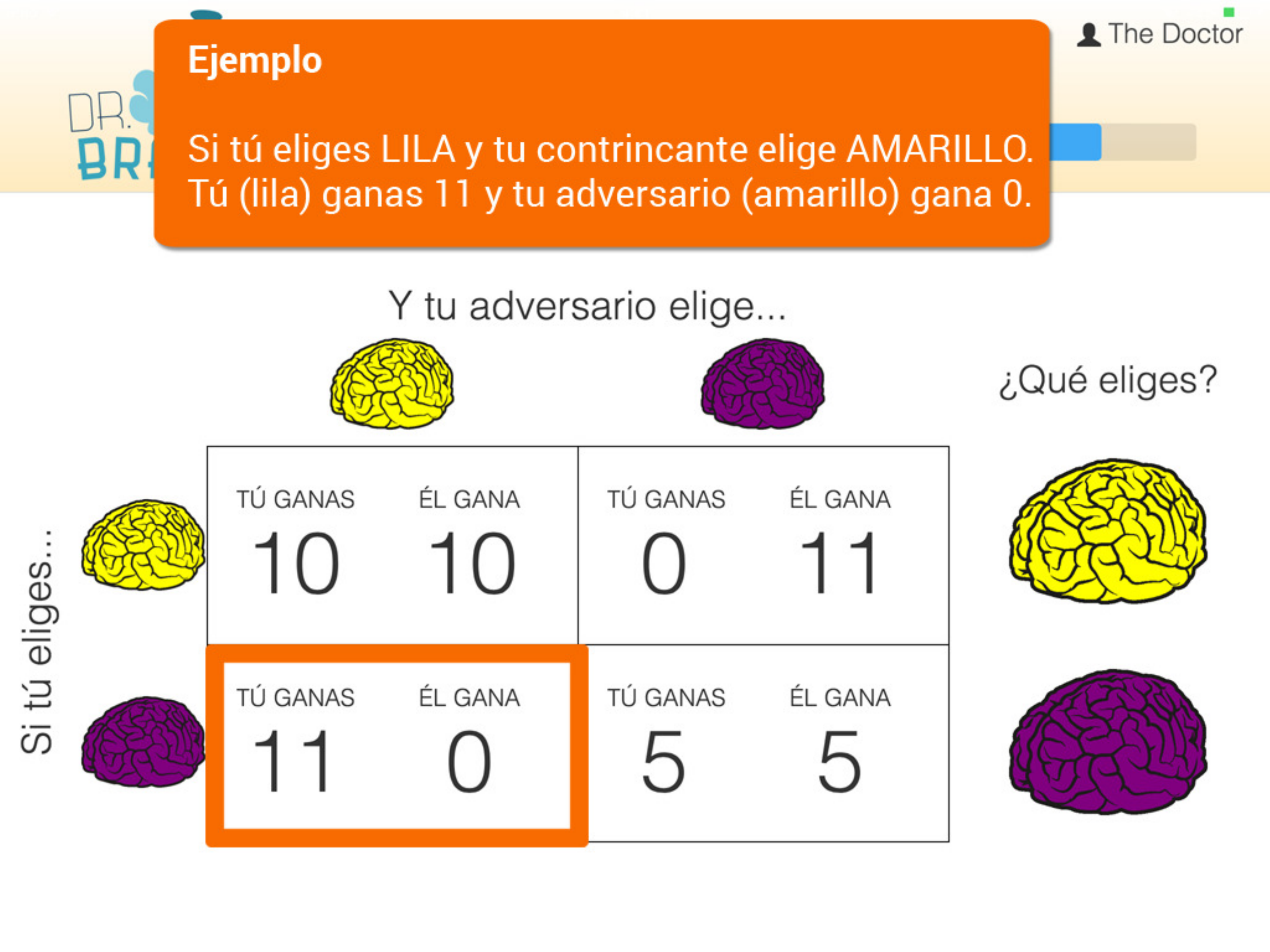} \\

(j)
\includegraphics[width=0.3\textwidth]{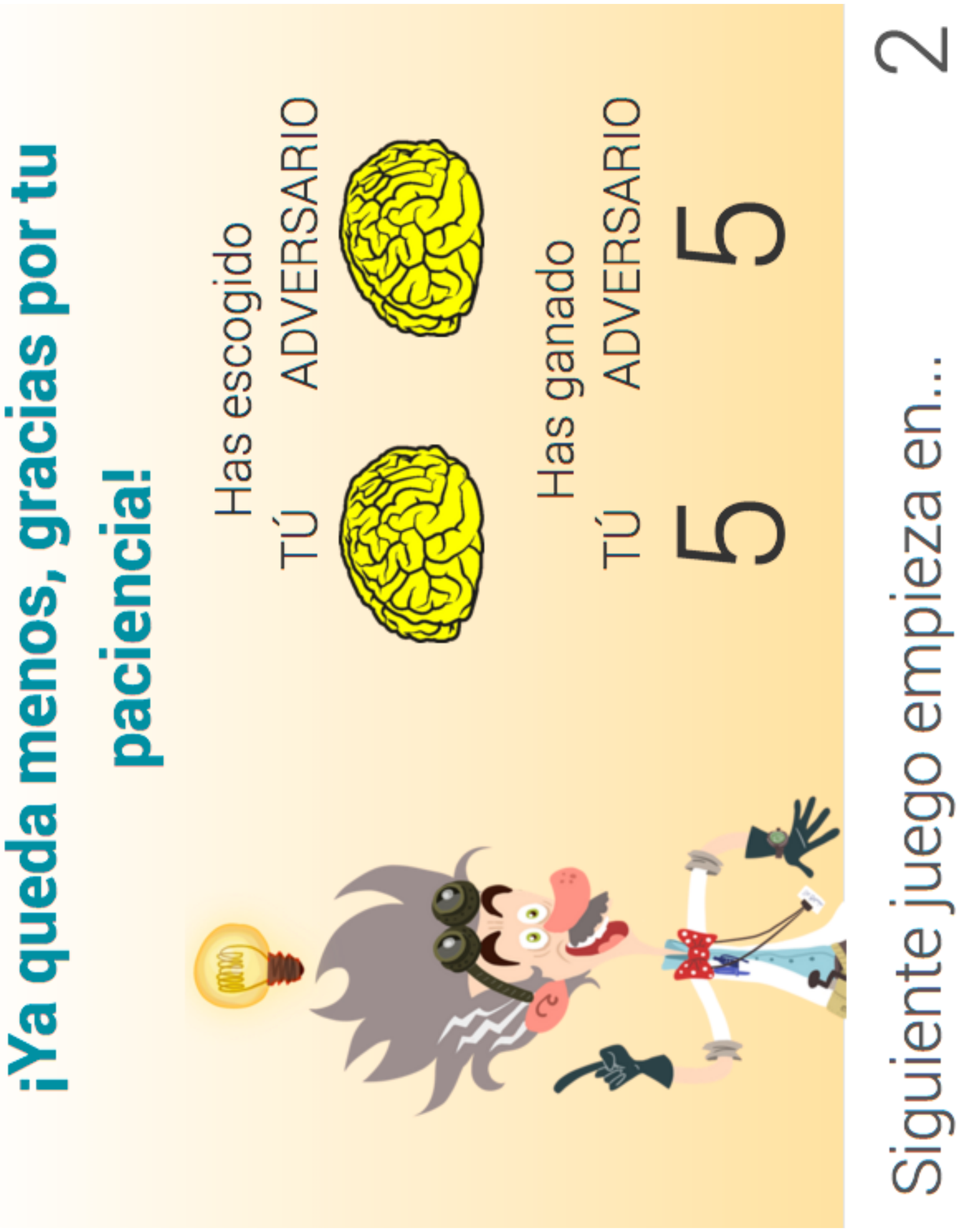} 

\end{array}$
\end{center}
\caption{Screenshots of the tutorial shown to participants before starting the experiment, and feedback screen after a typical round of the game. See text for translation.}\label{fig:tutorial_1}
\end{figure*}

\clearpage

\begin{figure*}[h]    
\begin{center}$
\begin{array}{lll}
\includegraphics[width=0.25\textwidth,
angle=-90]{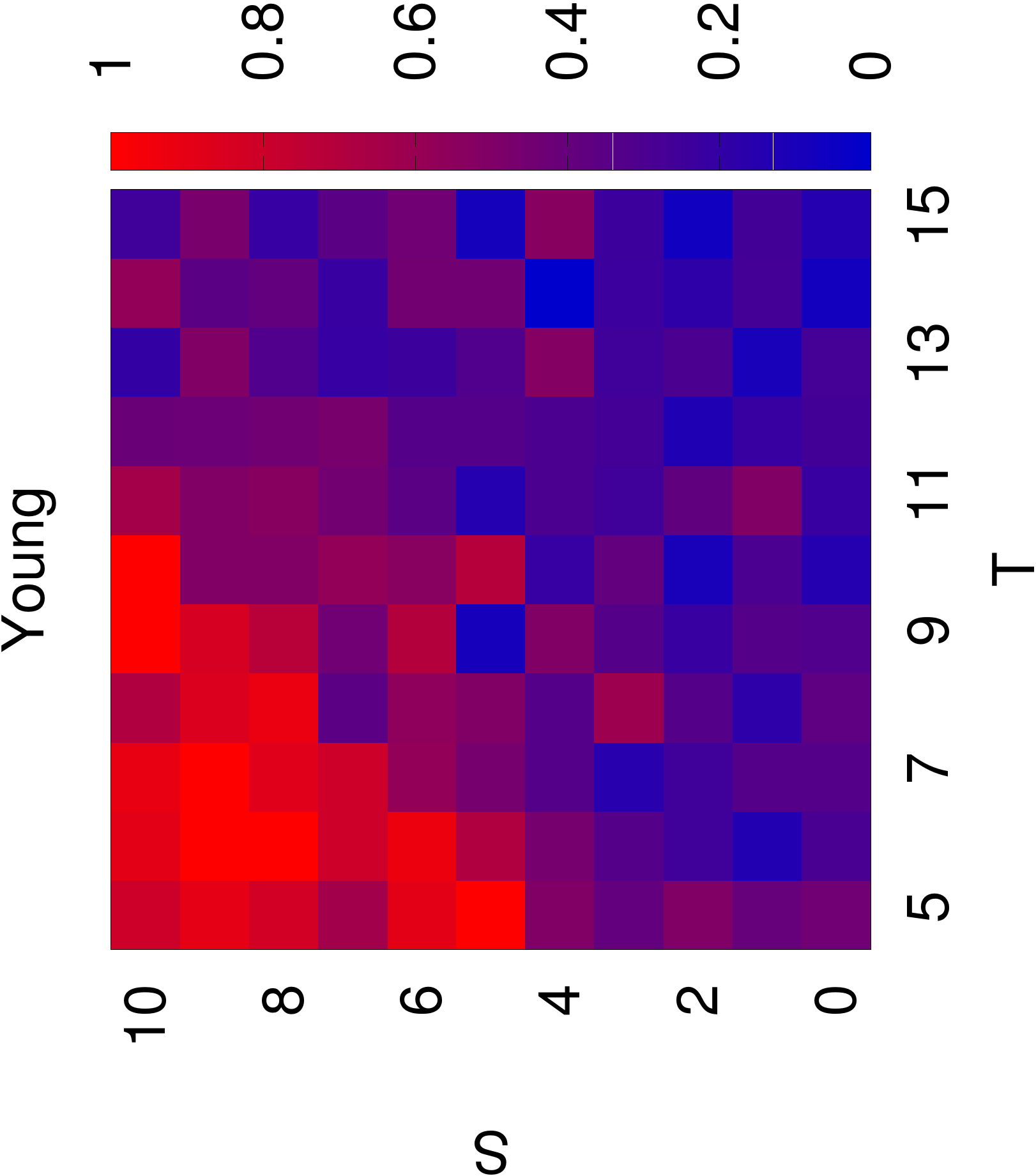} 
\includegraphics[width=0.25\textwidth, angle=-90]{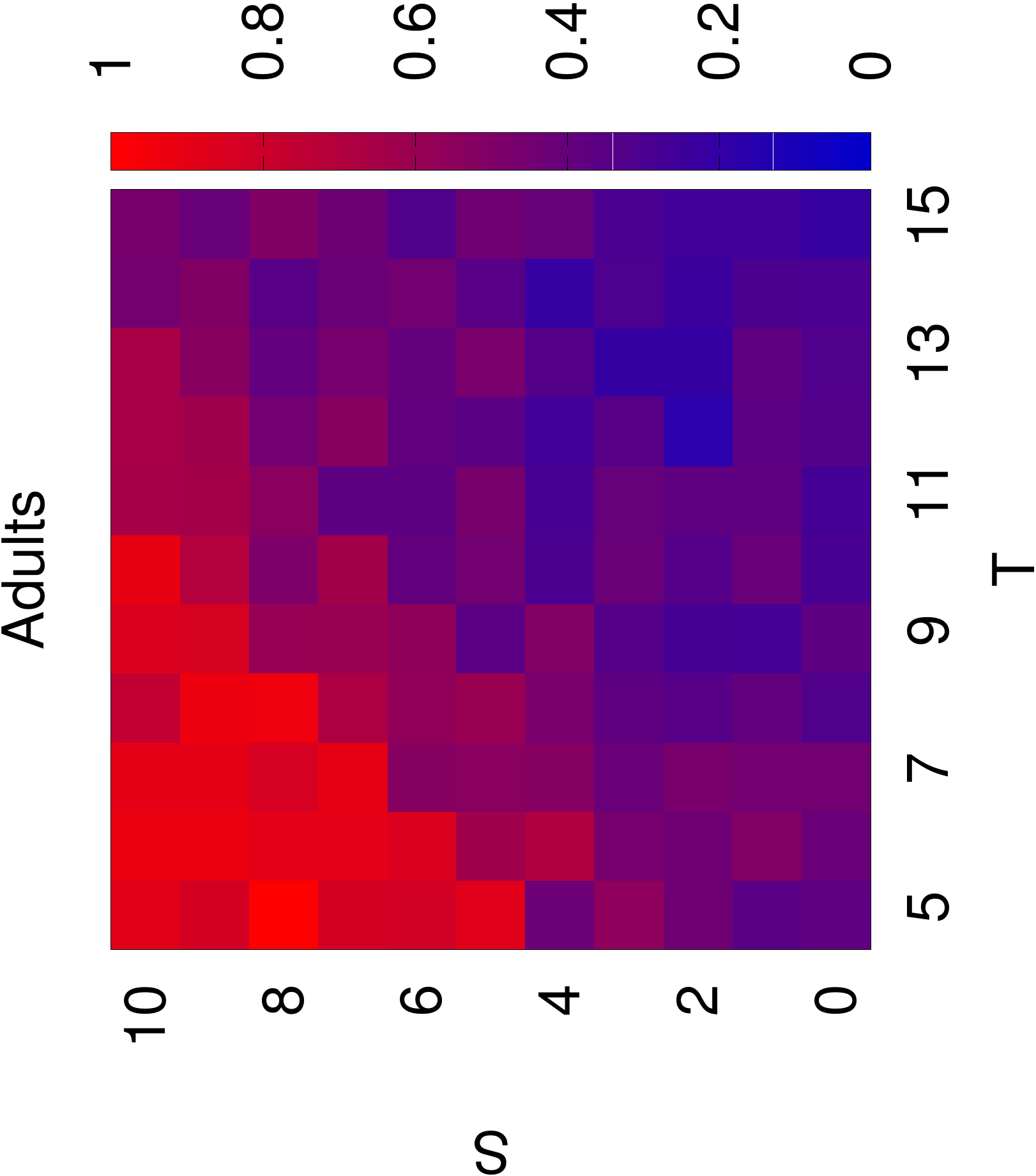} 
\includegraphics[width=0.25\textwidth, angle=-90]{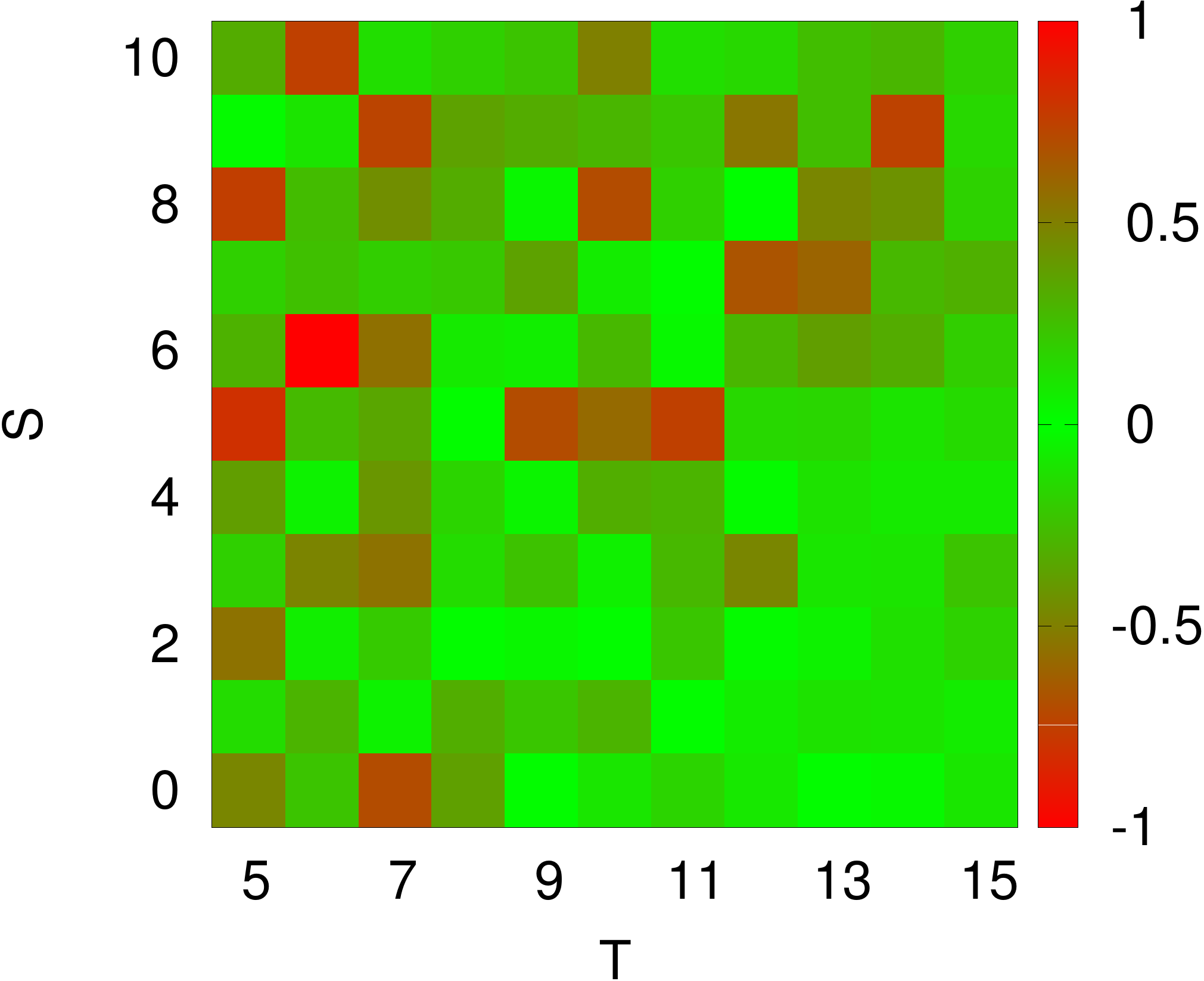} 
\end{array}$
\end{center}
\caption{Fraction of cooperative actions for young ($\leq 15$ years old) and adult players ($>16$ years old), and relative difference between the two heatmaps: (young-adults)/adults.}\label{fig:heatmaps_young_old}
\end{figure*}

\begin{figure*}[h]  
\begin{center}$
\begin{array}{cc}
\includegraphics[width=0.25\textwidth, angle=-90]{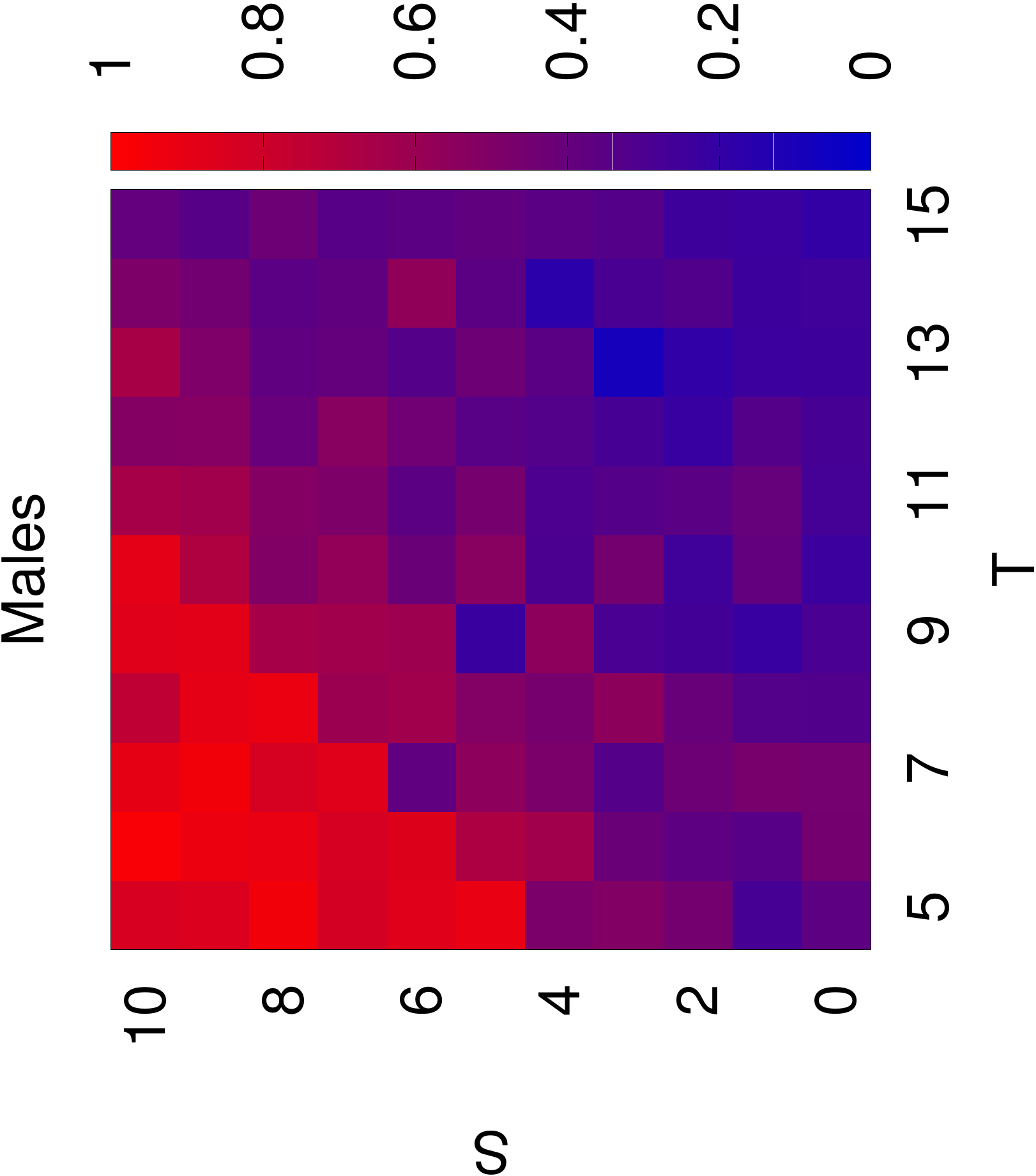} 
\includegraphics[width=0.25\textwidth, angle=-90]{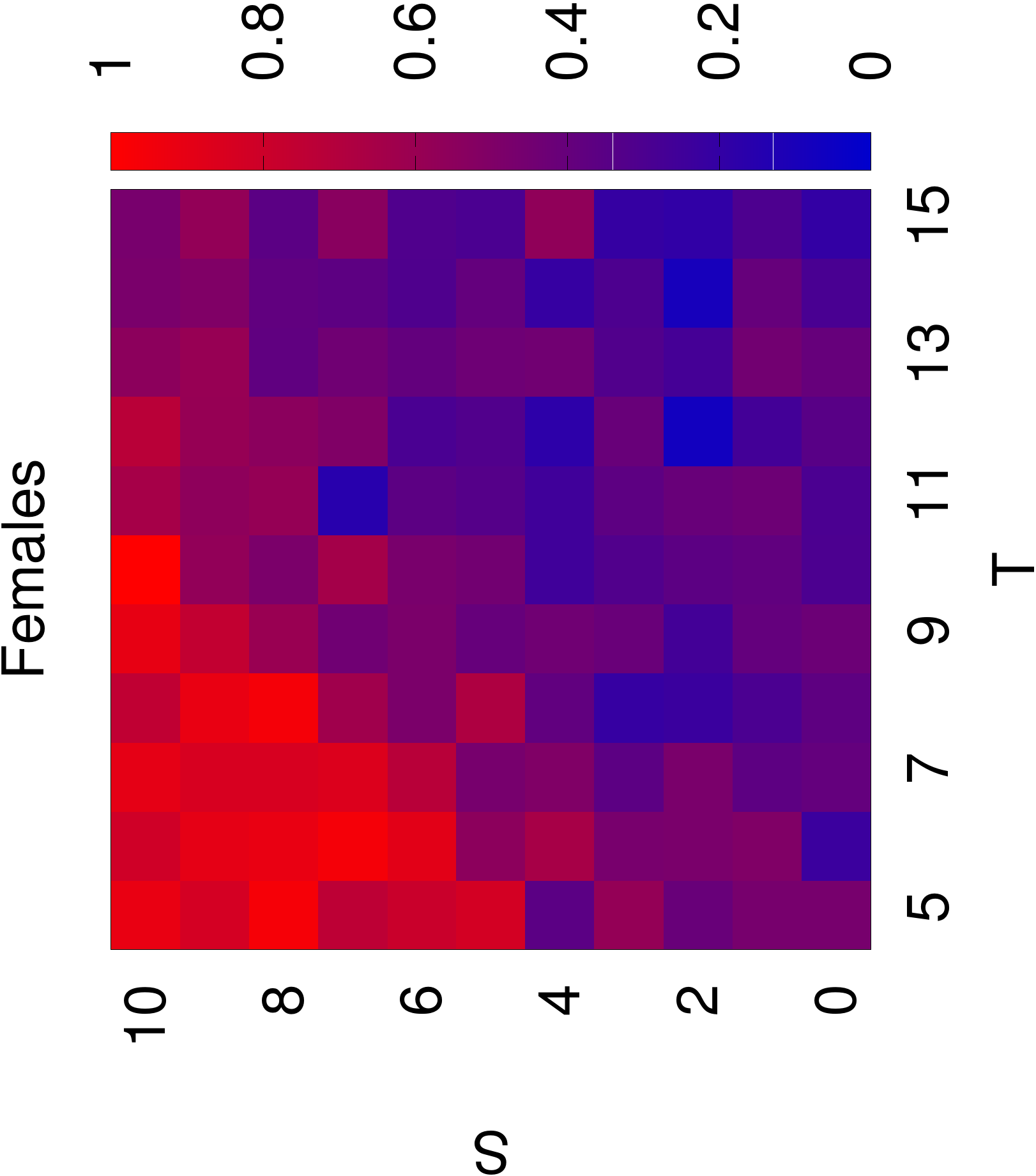}
\includegraphics[width=0.25\textwidth, angle=-90]{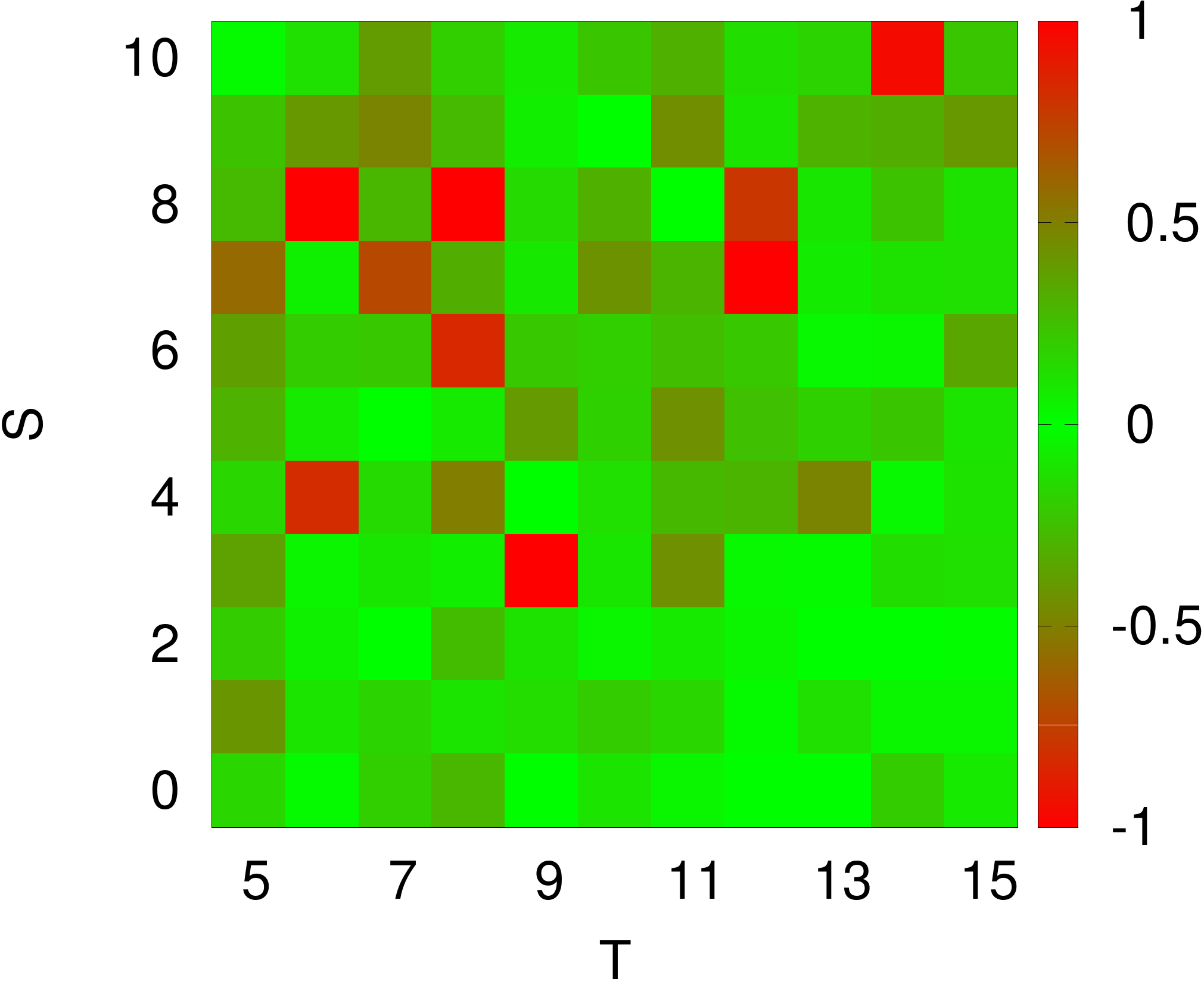} 
\end{array}$
\end{center}
\caption{Fraction of cooperative actions for males and females separately, and relative difference between the two heatmaps: (males-females)/females.}\label{fig:heatmaps_male_female}
\end{figure*}

\begin{figure*}[h]   
\begin{center}$
\begin{array}{c}
\includegraphics[width=0.25\textwidth, angle=-90]{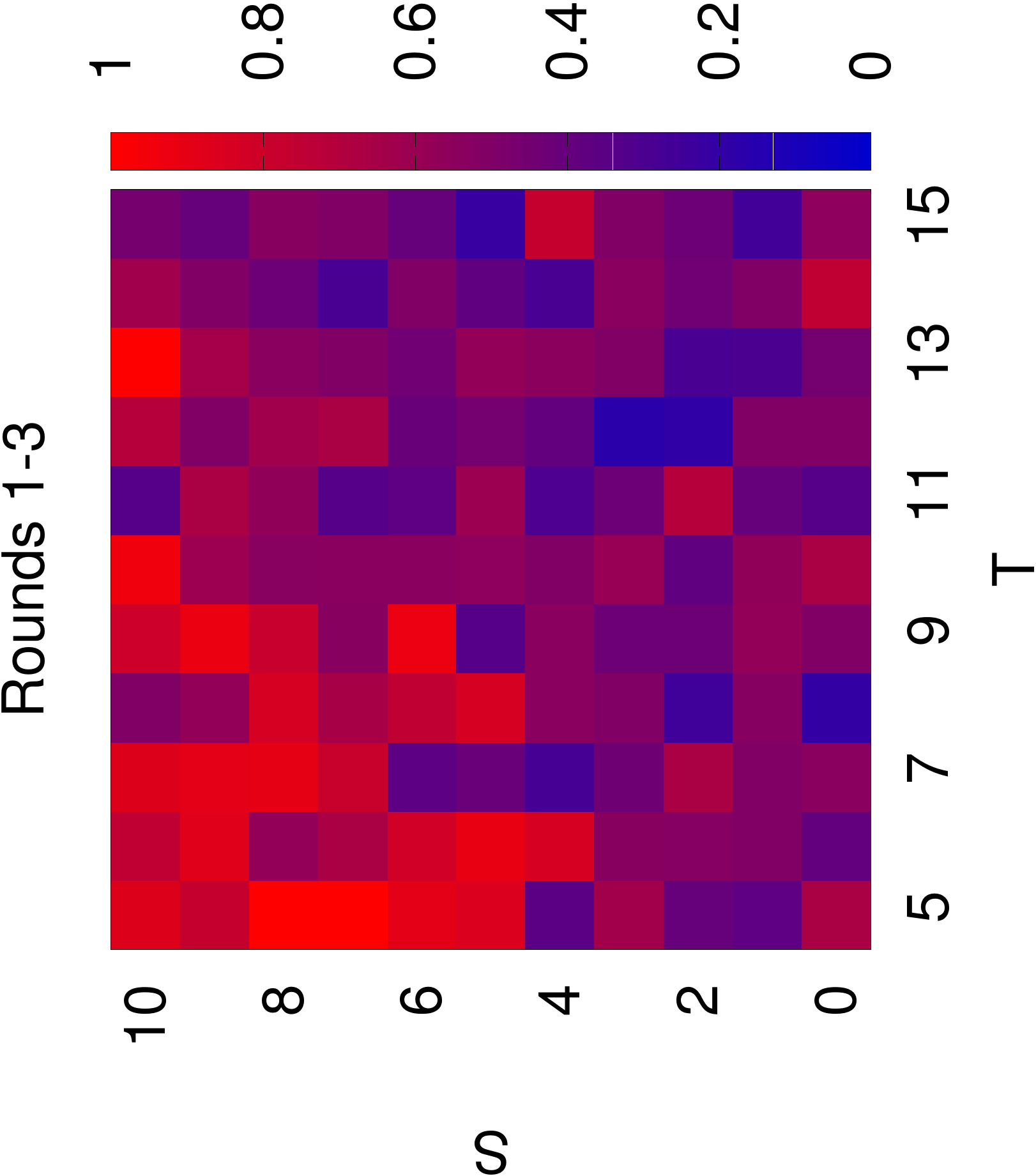} 
\includegraphics[width=0.25\textwidth, angle=-90]{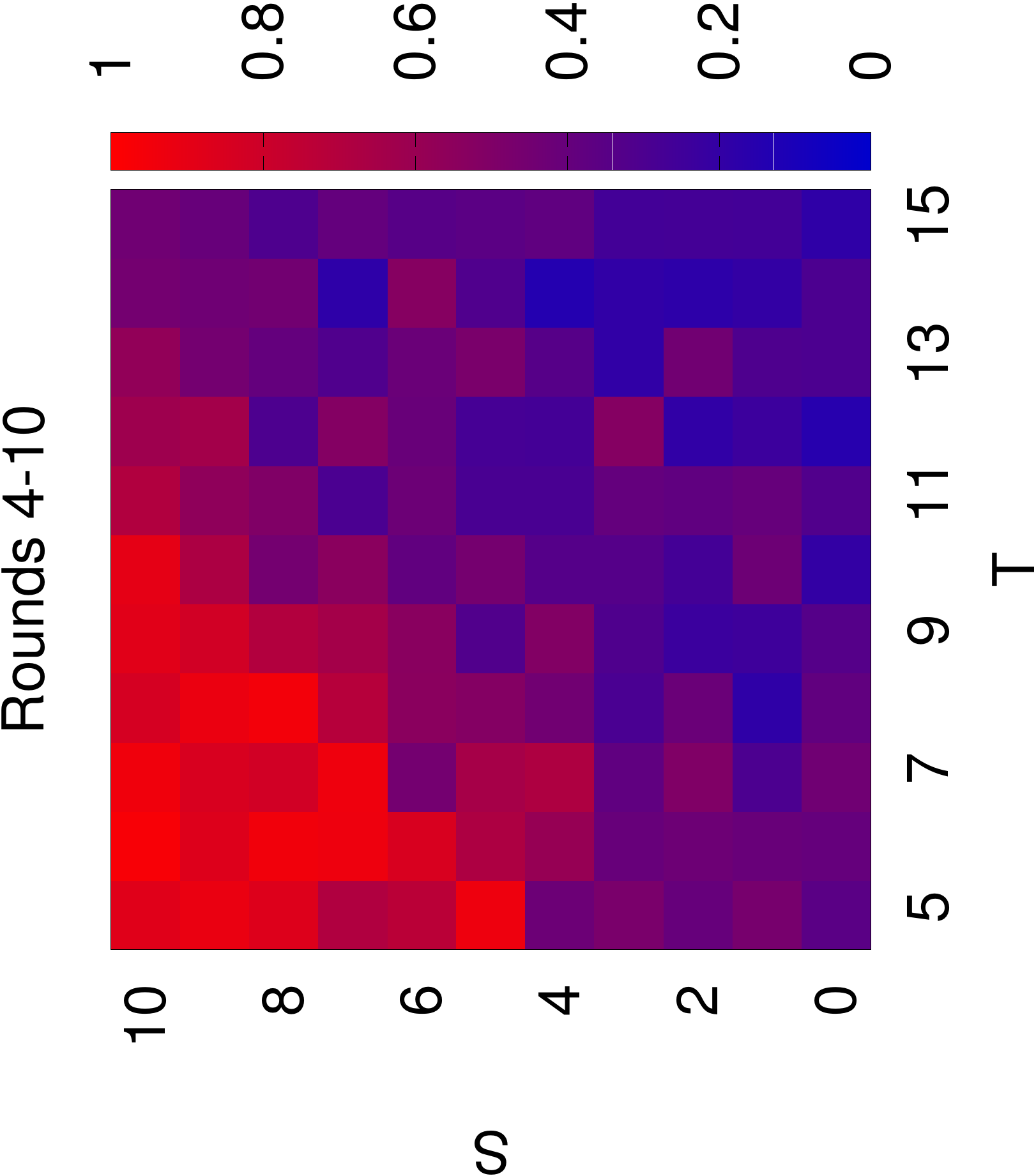} 
\includegraphics[width=0.25\textwidth, angle=-90]{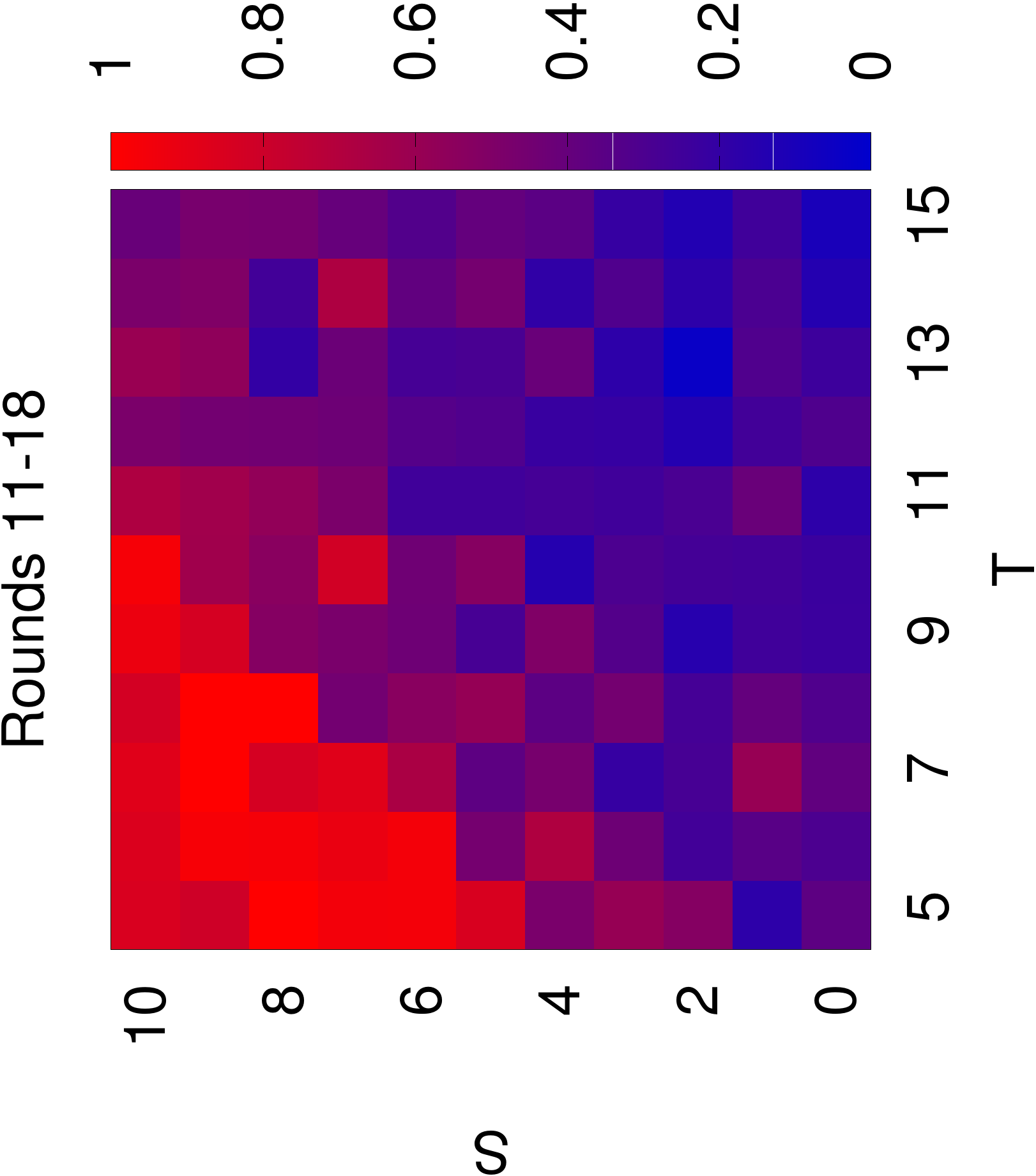} 
\end{array}$
\end{center}
\caption{Fraction of cooperative actions separating by round number: for the first 1 to 3 rounds, 4 to 10 and last 11 to 18 rounds.}\label{fig:heatmaps_by_round}
\end{figure*}

\clearpage

\begin{figure*}[h]   
\begin{center}$
\begin{array}{c}
\includegraphics[width=0.25\textwidth, angle=-90]{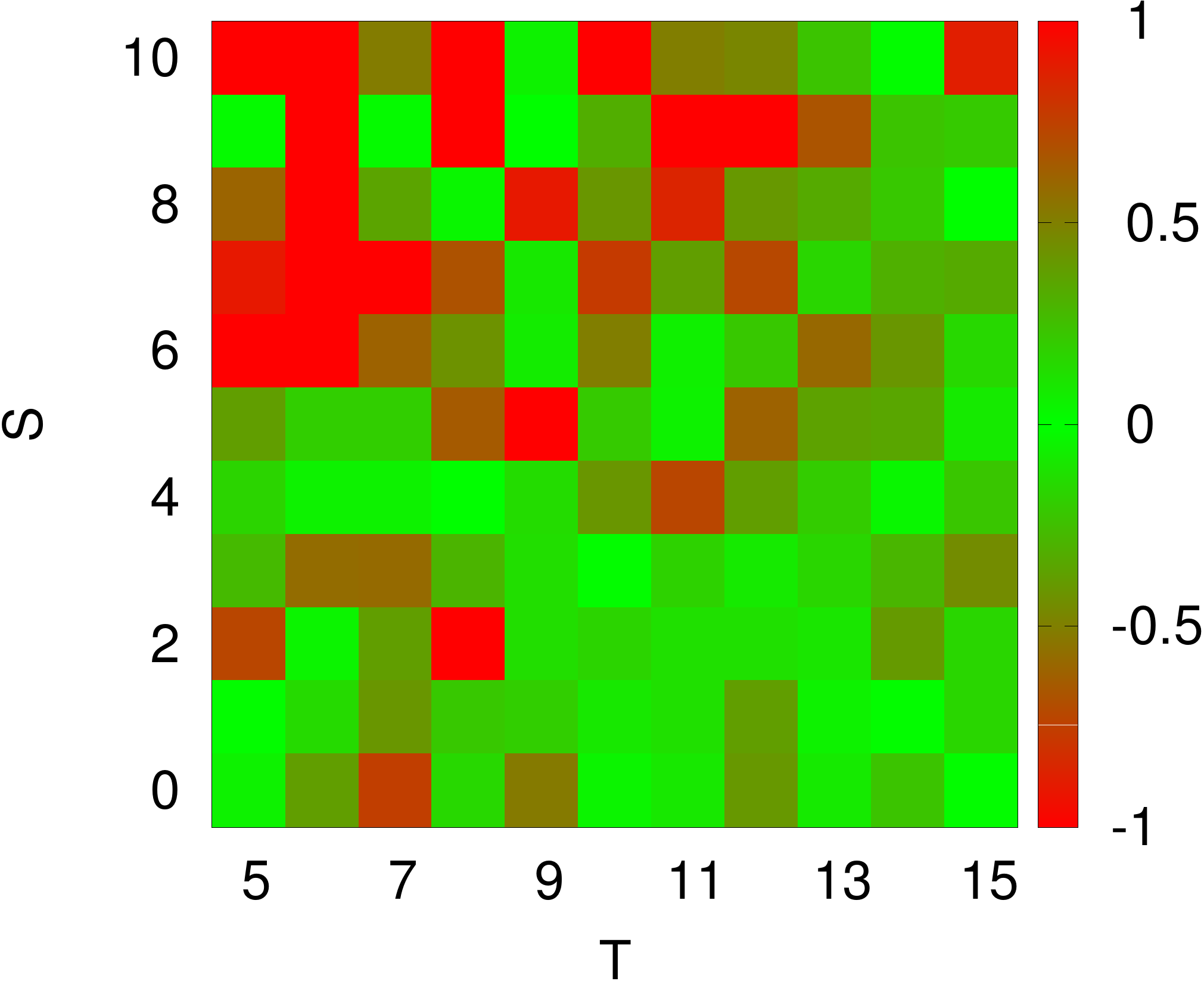} 
\includegraphics[width=0.25\textwidth, angle=-90]{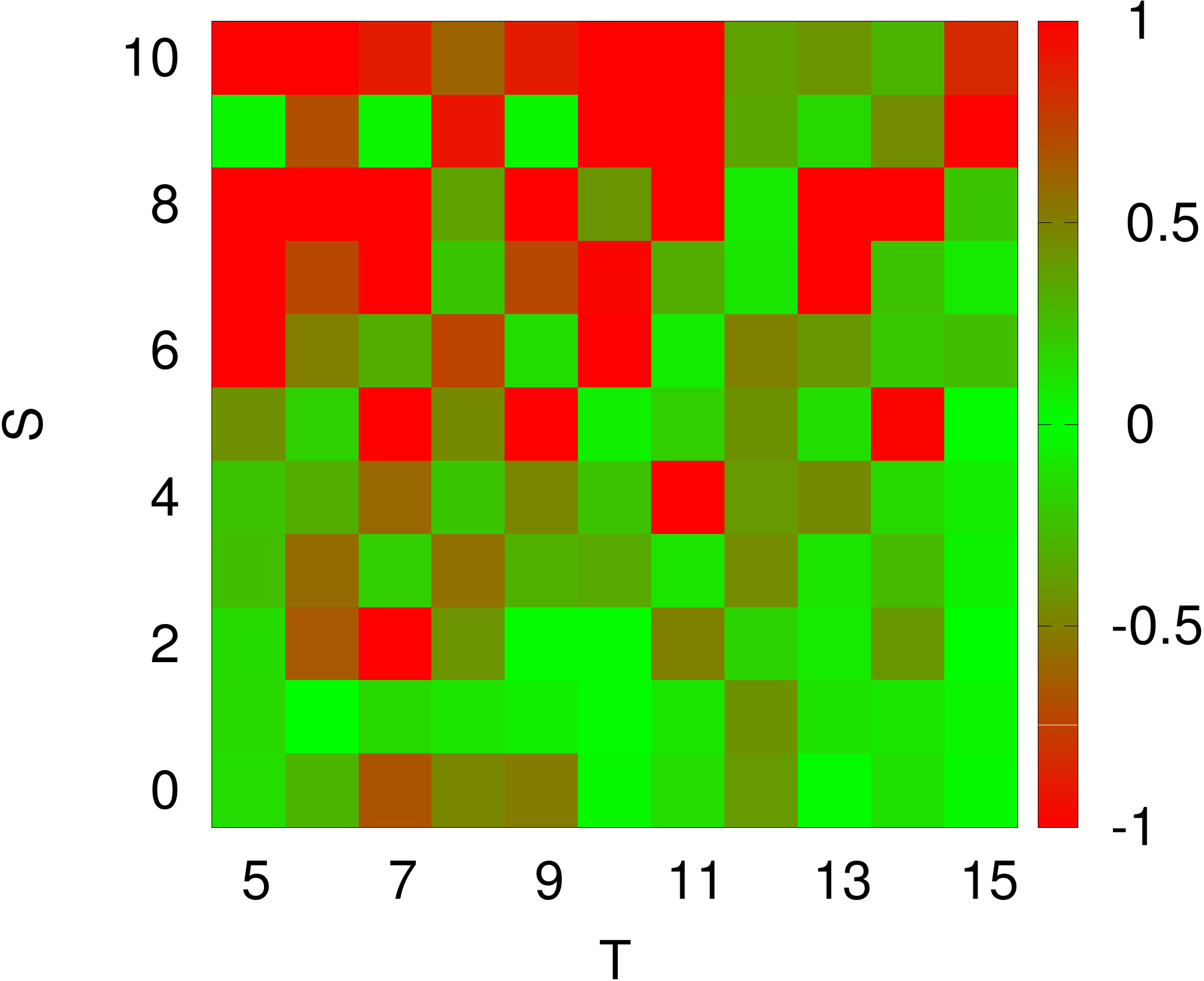} 
\includegraphics[width=0.25\textwidth, angle=-90]{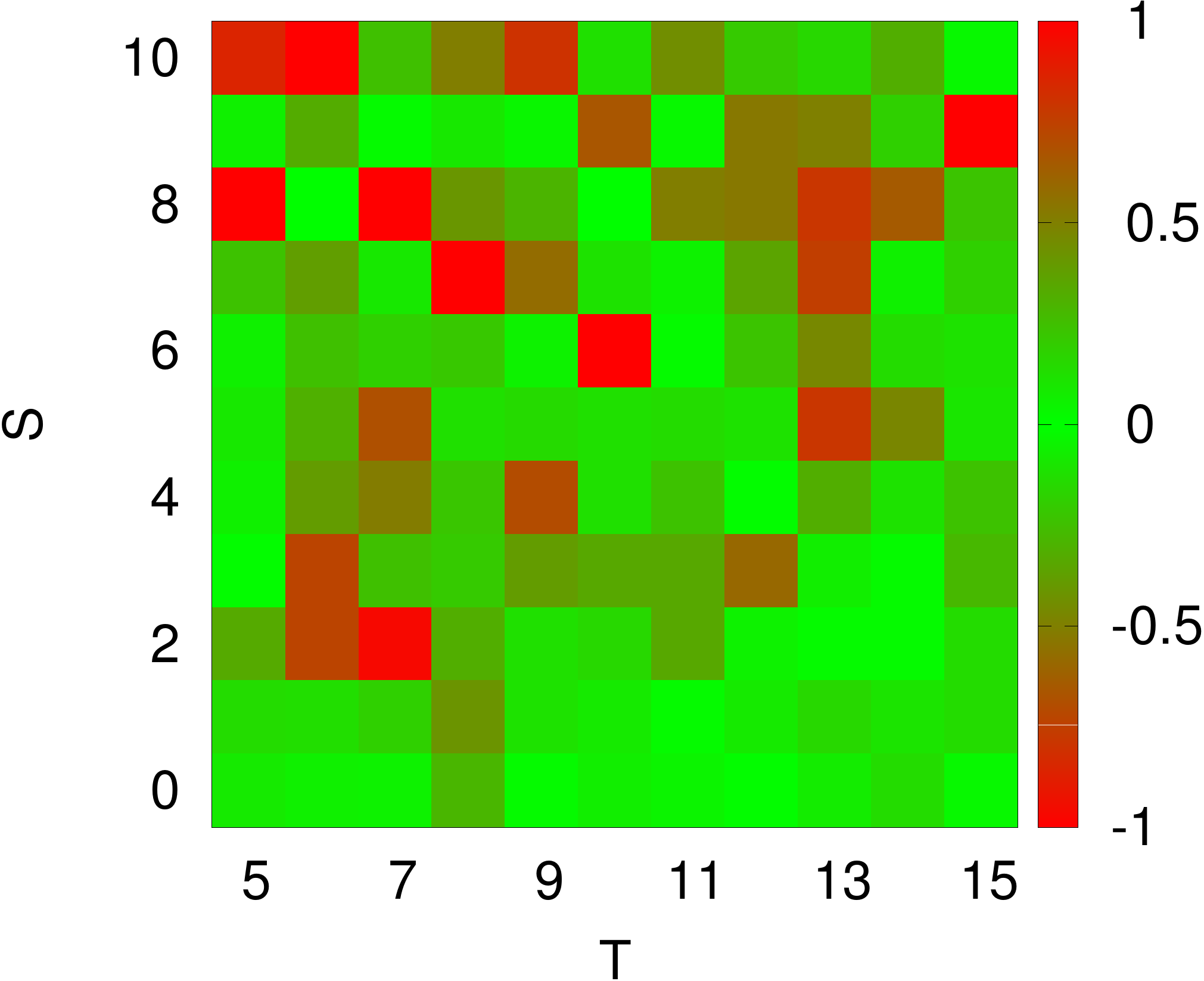} 
\end{array}$
\end{center}
\caption{Relative difference in the fraction of cooperation heatmaps between groups of rounds. Left: (rounds 1 to 3 - rounds 4 to 10)/(rounds 4 to 10); Center: (rounds 1 to 3 - rounds 11 to 18)/(rounds 11 to 18) ; Right: (rounds 4 to 10 - rounds 11 to 18)/(rounds 11 to 18).
}\label{fig:heatmaps_by_round_relative_diff}
\end{figure*}

\begin{figure}[h]  
\begin{center}$
\begin{array}{cc}
\includegraphics[width=0.65\textwidth, angle=-90]{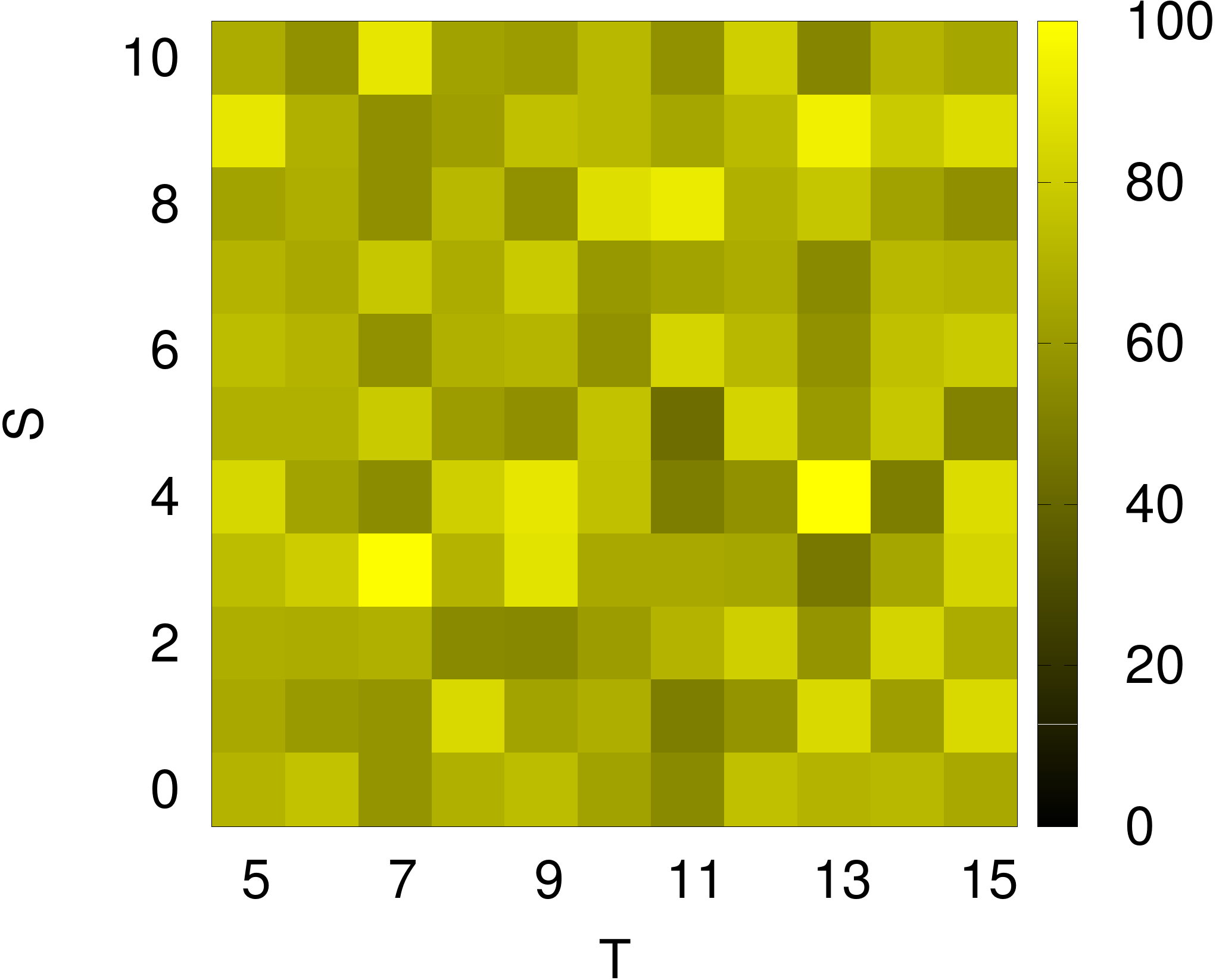} 
\end{array}$
\end{center}
\caption{Total number of actions in each point of the $(T,S)$-plane, for all $541$ participants in the experiment (the total number of game actions in the experiment adds up to $8,366$).}\label{fig:Num_actions_cooperation_map}
\end{figure}

\begin{figure}[h]   
\begin{center}$
\begin{array}{cc}
\includegraphics[width=0.65\textwidth, angle=-90]{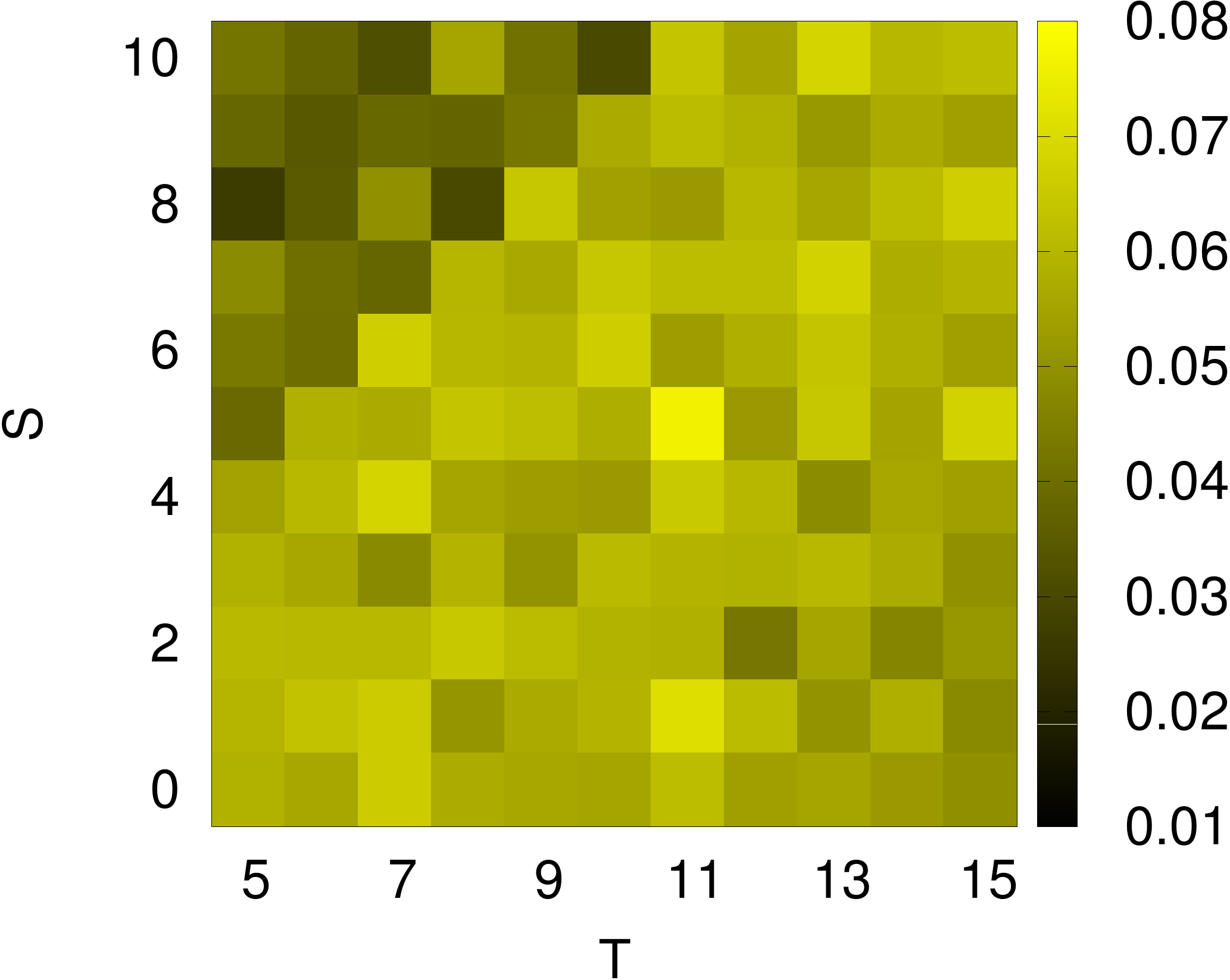}
\end{array}$
\end{center}
\caption{Standard Error of the Mean (SEM) fraction of cooperative actions in each point of the $(T,S)$-plane, for all the participants in the experiment. The HG regions leads to lower SEM of cooperation and that was expected given that two important types of phenotypes predict cooperation. To get a 95$\%$ Confidence Interval errors bars should be multiplied by 1.96.}\label{fig:SEM_cooperation_map}
\end{figure}

\begin{figure*}[h]  
\begin{center}$
\begin{array}{cc}
\includegraphics[width=0.45\textwidth]{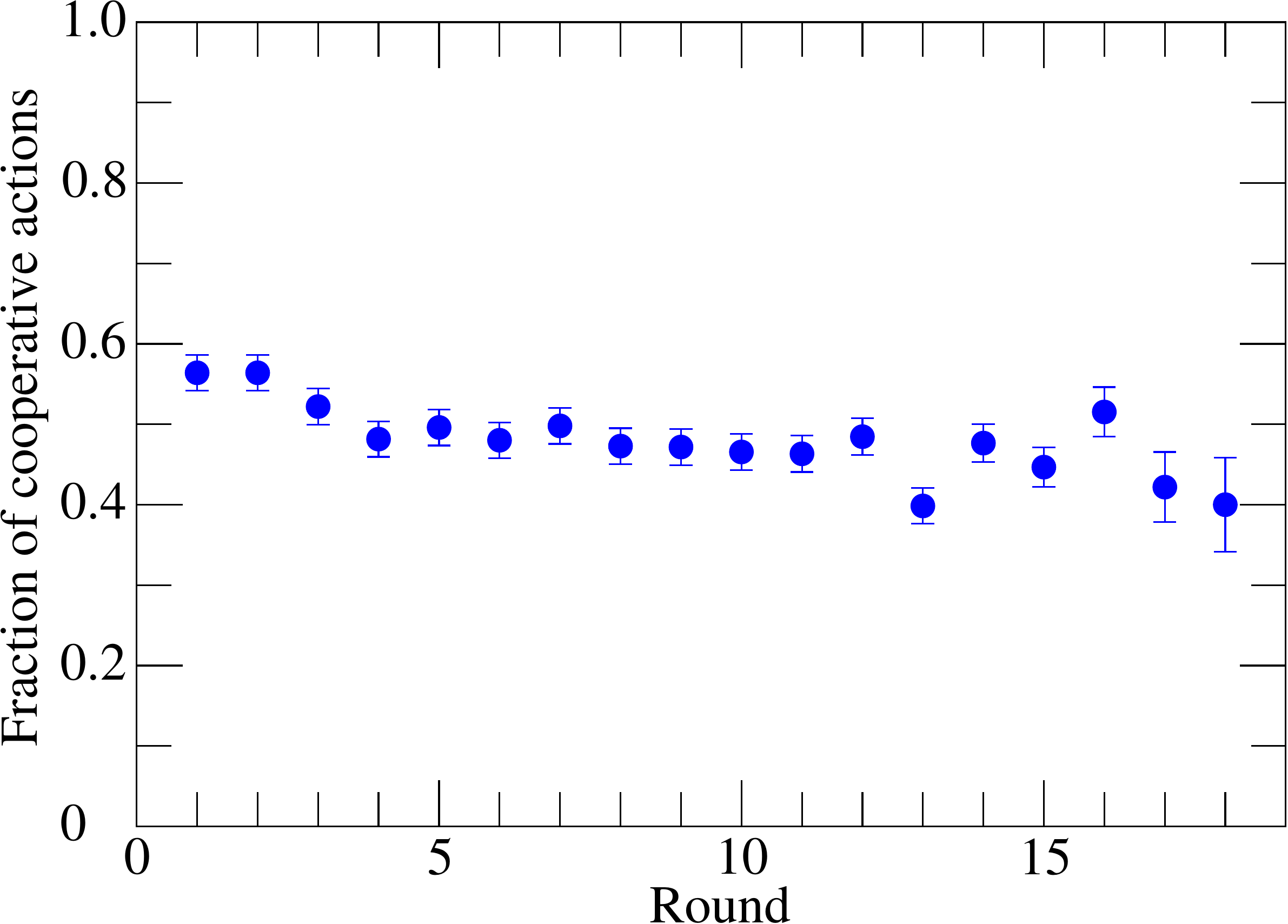} \hspace{8mm}
\includegraphics[width=0.4\textwidth]{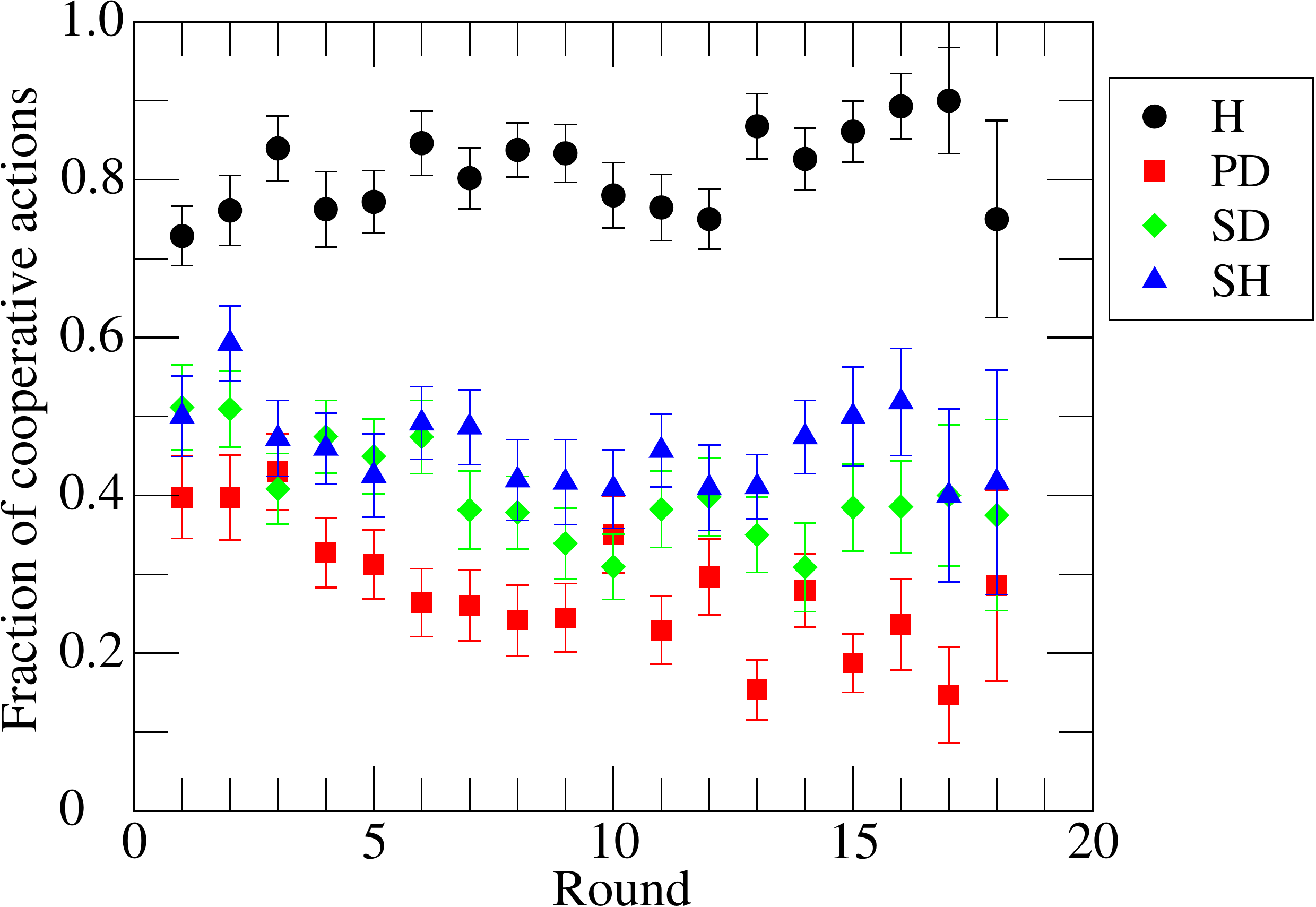} 
\end{array}$
\end{center}
\caption{Average fraction of cooperative actions (and Standard Error of the Mean) among the population as a function of the round number overall (left) and separating the actions by game (right).}\label{fig:Coop_vs_time}
\end{figure*}

\begin{figure}[h]  
\begin{center}$
\begin{array}{cc}
\includegraphics[width=0.75\textwidth]{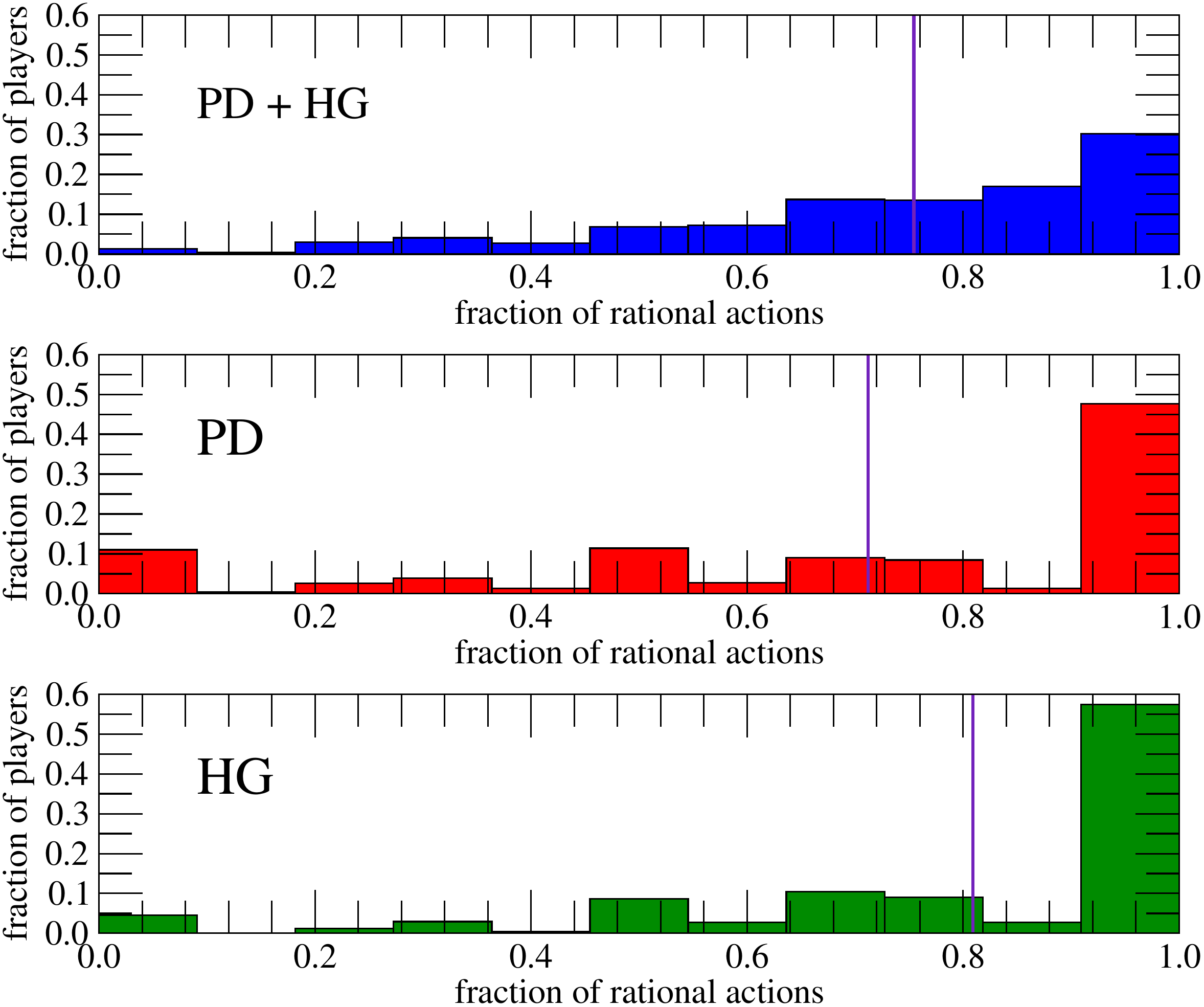} 
\end{array}$
\end{center}
\caption{Distribution of fraction of rational actions among the $541$ subjects of our experiment, when considering only their actions in the Harmony game (HG), or the Prisoner's Dilemma (PD), or both together. The purple line indicates the mean value. }\label{fig:hist_rationality}
\end{figure}

\begin{figure}[h]   
\begin{center}$
\begin{array}{cc}
\includegraphics[width=0.75\textwidth]{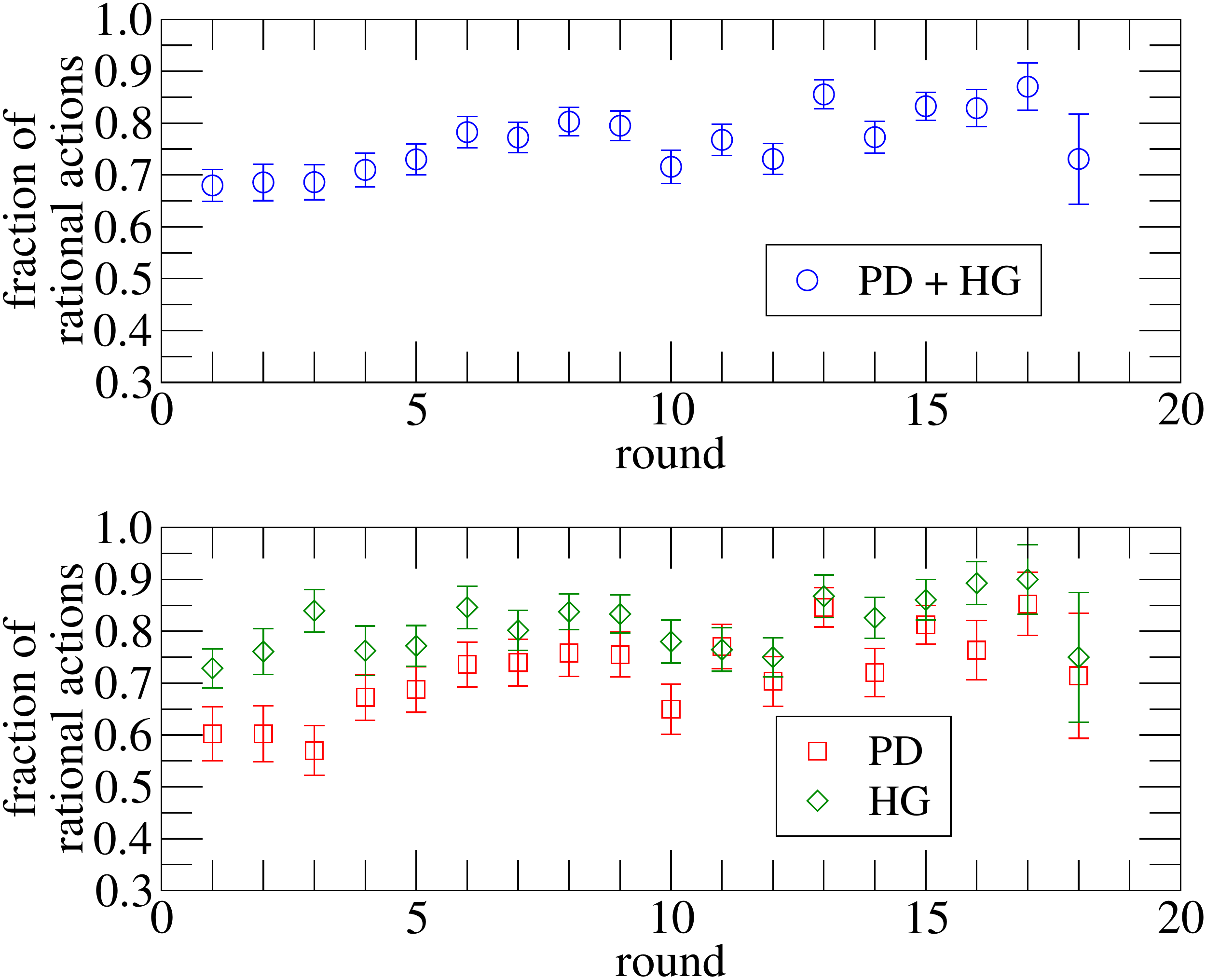} 
\end{array}$
\end{center}
\vspace{2mm}
\caption{Fraction of rational actions as a function of the round number for the 541 subjects, defined by their actions in the Prisoner's Dilemma game (PD) and Harmony game (HG) together (top), and independently (bottom). The bars correspond to the Standard Error of the Mean.} \label{fig:time_evol_rationality}
\end{figure}

\begin{figure}[h]   
\begin{center}$
\begin{array}{cc}
\includegraphics[width=0.95\textwidth]{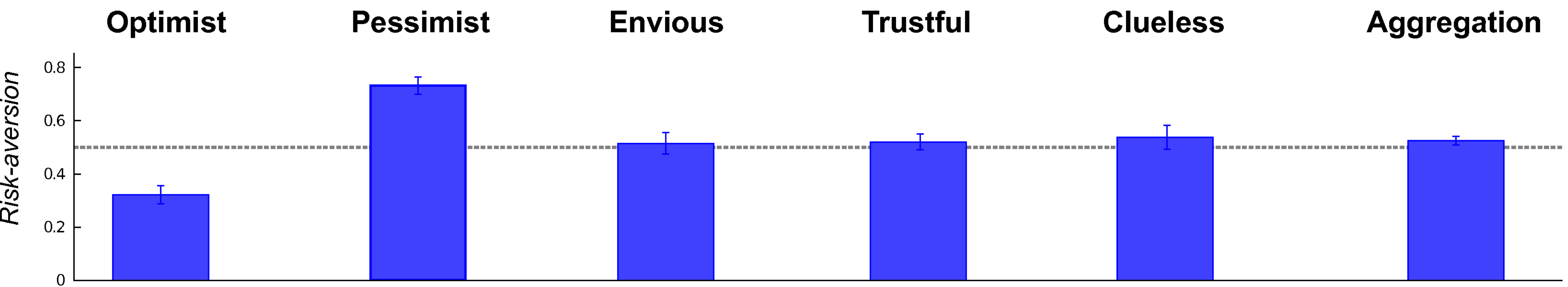} 
\end{array}$
\end{center}
\vspace{2mm}
\caption{Values of risk-aversion averaged over the subjects in each phenotype. The phenotypes of Optimist and Pessimist show  significantly lower and higher values than random expectation, respectively. Error bars indicate $95\%$ Confidence Intervals.} \label{fig:risk_aversion}
\end{figure}

\begin{figure}[h]  
\begin{center}$
\begin{array}{cc}
\includegraphics[width=0.75\textwidth]{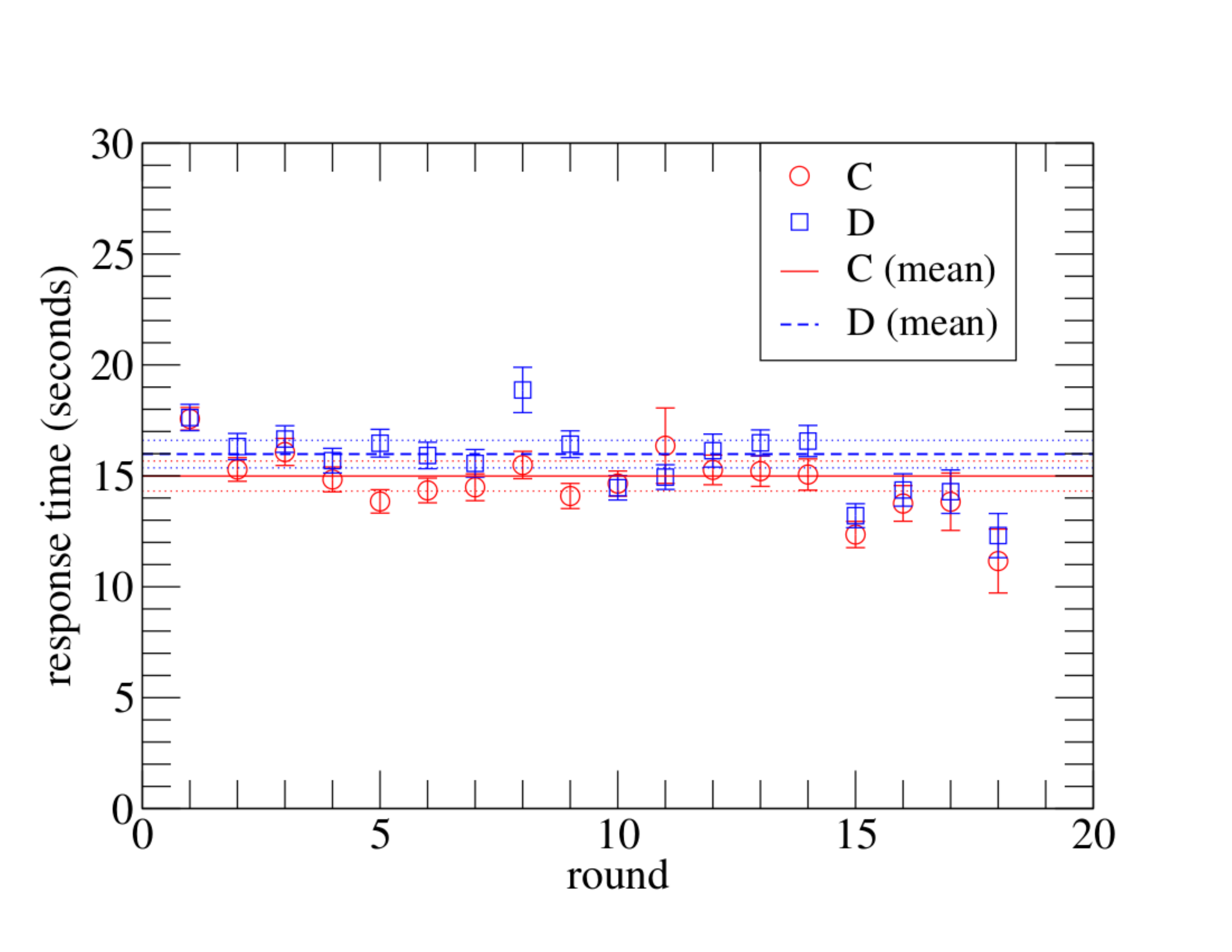} 
\end{array}$
\end{center}
\caption{Average response times (and Standard Error of the Mean) as a function of the round number, for all the participants in the experiment, and separating the actions into cooperation ($C$) or defection ($D$).}\label{fig:response_time}
\end{figure}

\begin{figure}[h]   
\begin{center}$
\begin{array}{cc}
\includegraphics[width=0.65\textwidth, angle=-90]{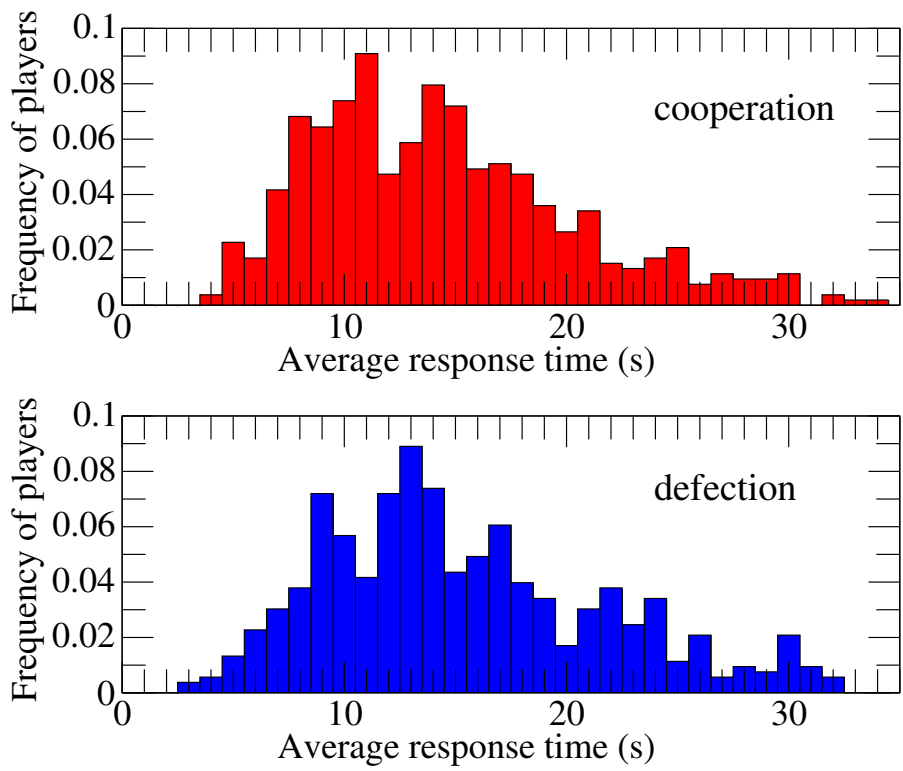} 
\end{array}$
\end{center}
\caption{Distributions of response times for all the participants in the experiment, and separating the actions into cooperation (top) and defection (bottom).}\label{fig:response_time_hist}
\end{figure}

\begin{figure}[h] 
\begin{center}
\includegraphics[trim={0 0 0 0},clip,width=0.75\textwidth]{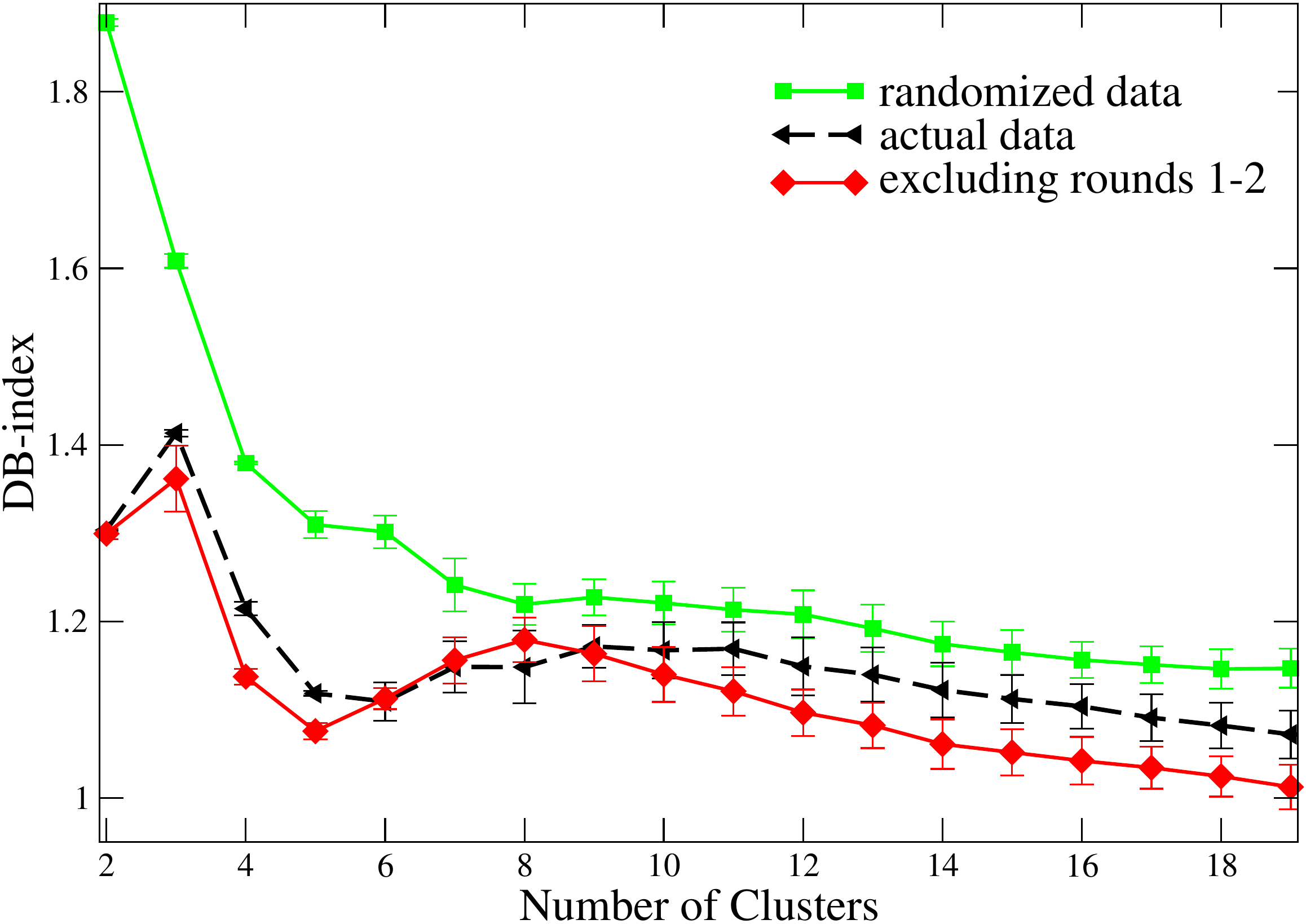}
\end{center}
\caption{Testing the robustness of the results from the $K$-means algorithm. We present the average value of the DB-index over 200 independent runs of the algorithm on the data, as a function of the number of clusters (black). The optimum number of clusters is 5 (we note that, although a 6-cluster partition is also comparably good, the Standard Deviation (SD) is larger in that case, indicating less stability across different runs). We also show the results for the case of a randomization of the data (green). In this case, we observe that there is no local minimum, indicating a lack of cluster structure. Finally, we observe that when excluding the first two choices of every subject in our experiment (to account for excessive noise due to lack of experience), the position of the optimum is located in a clearer way at 5.}\label{fig:robustness}
\end{figure}

\begin{figure}[h]  
\begin{center}$
\begin{array}{cc}
\includegraphics[width=0.75\textwidth]{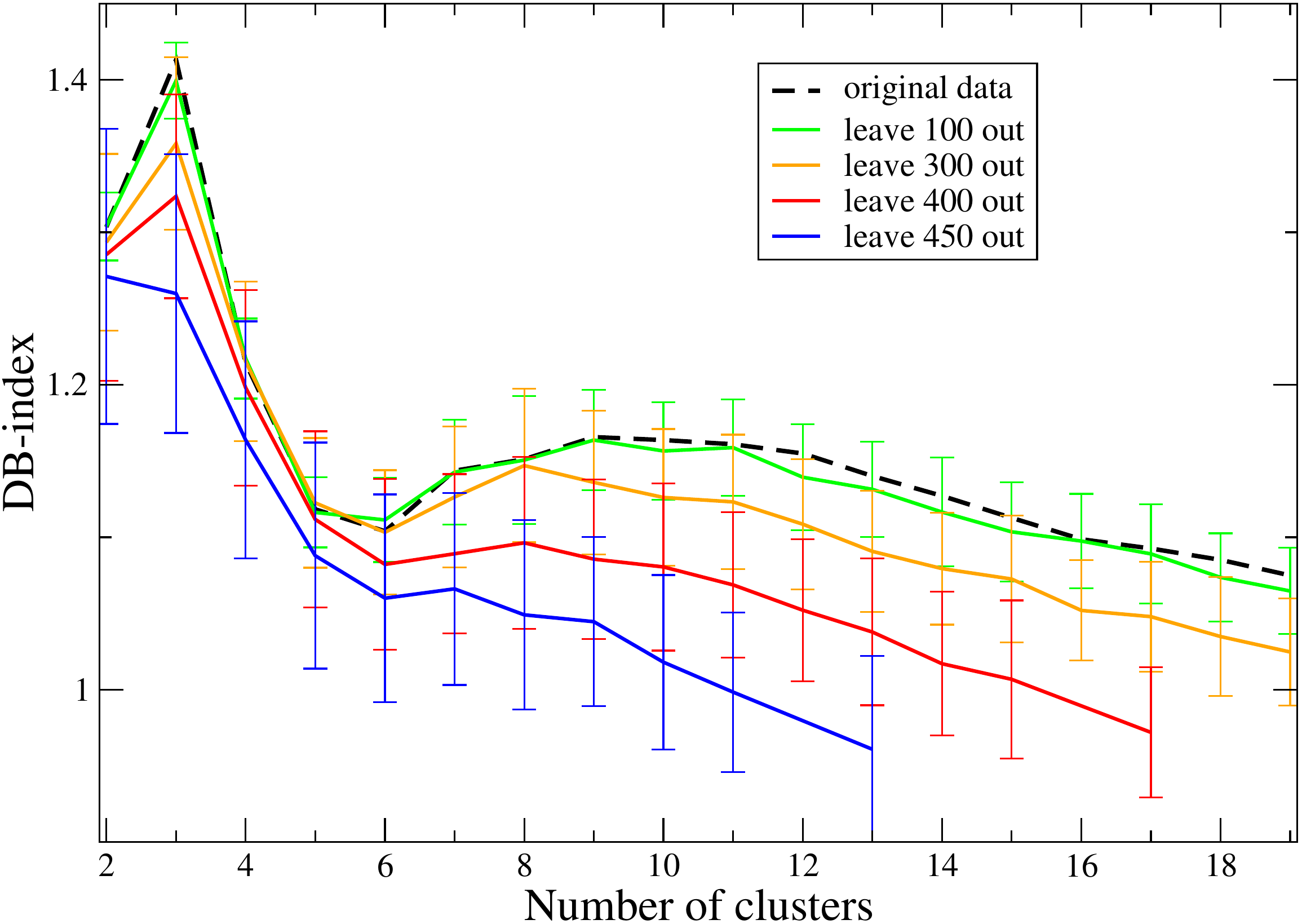}
\end{array}$
\end{center}
\caption{DB-index as a function of the number of clusters in the partition of our data (dashed black) as it compares to the equivalent results for different leave-p-out analyses. See text for details. The bars correspond to the Standard Deviation over the 200 independent realizations of the algorithm.} \label{fig:DBindex_leave_p_out}
\end{figure}

\begin{figure}[h]   
\begin{center}$
\begin{array}{cc}
\includegraphics[width=0.75\textwidth]{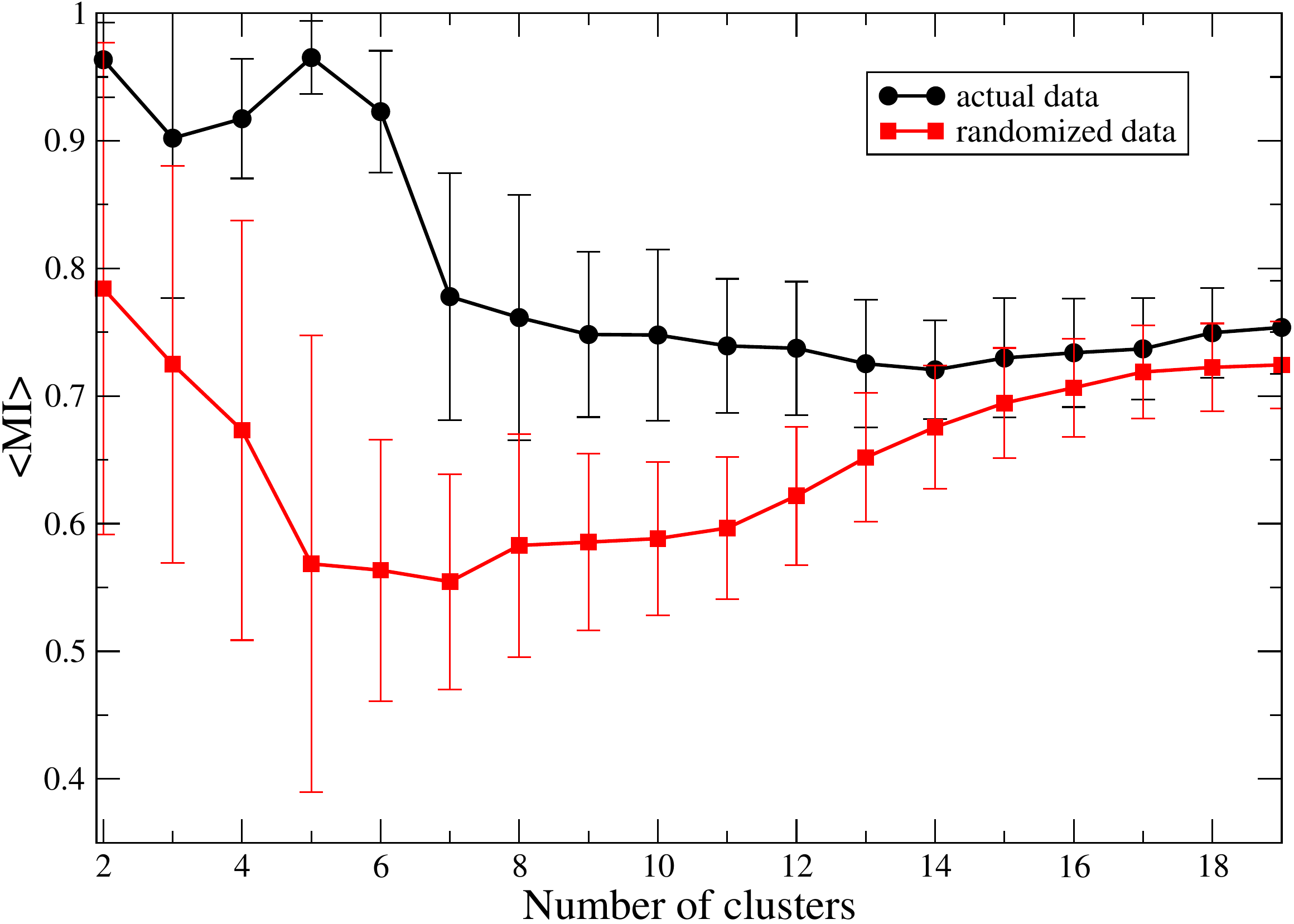}
\end{array}$
\end{center}
\caption{Average value for the Normalized Mutual Information Score, when doing pair-wise comparisons of the clustering schemes from $2,000$ independent runs of the $K$-means algorithm, both on the actual data, and on the randomized version of the data.} \label{fig:mutual_info}
\end{figure}

\begin{figure}[h]   
\begin{center}$
\begin{array}{cc}
\includegraphics[width=0.75\textwidth]{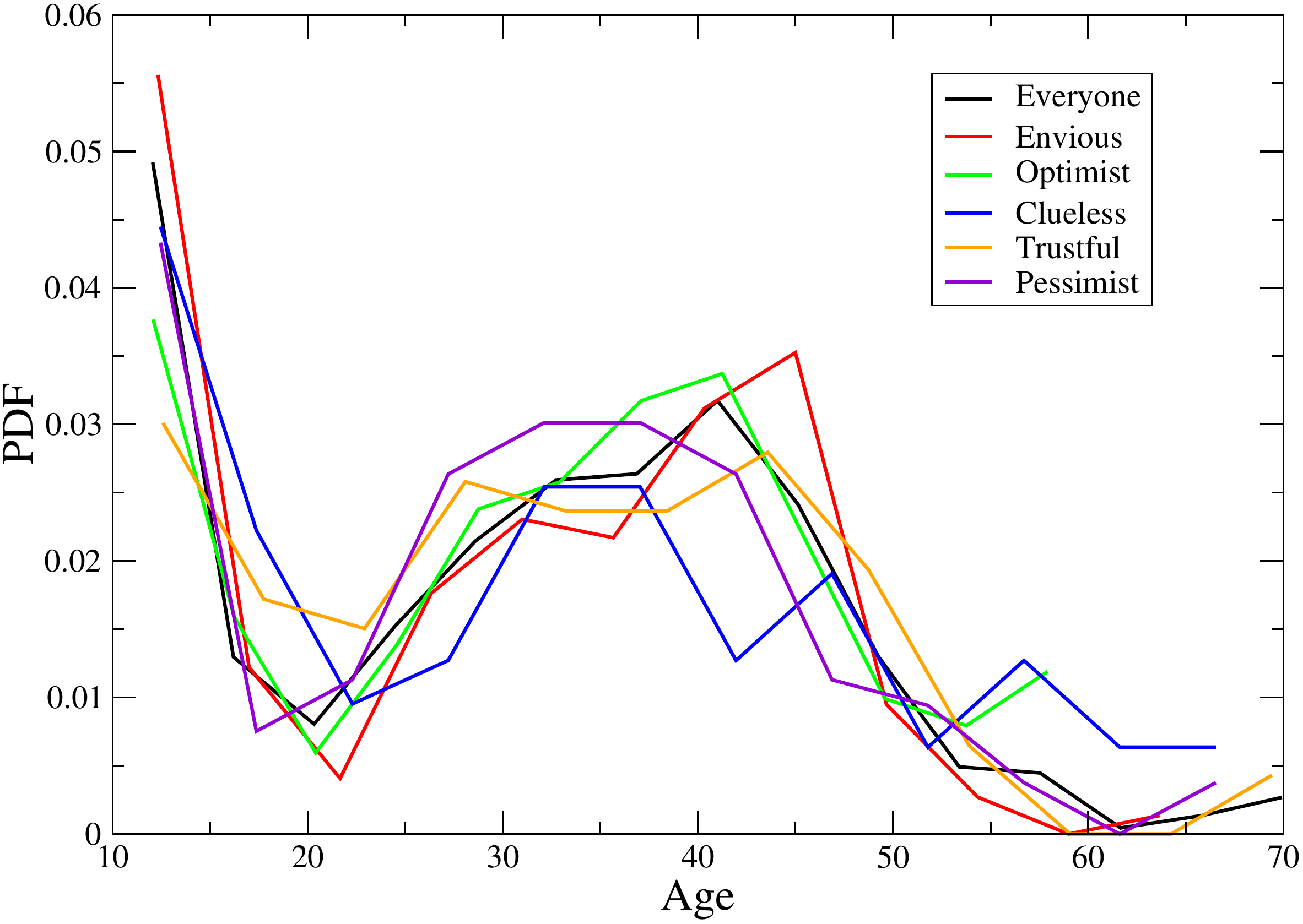}
\end{array}$
\end{center}
\caption{Age distribution for the different phenotypes, as it compares to the distribution of the whole population (black).} \label{fig:age_distrib_clusters}
\end{figure}

\begin{figure*}[h]   
\begin{center}$
\begin{array}{cc}
\includegraphics[width=0.95\textwidth]{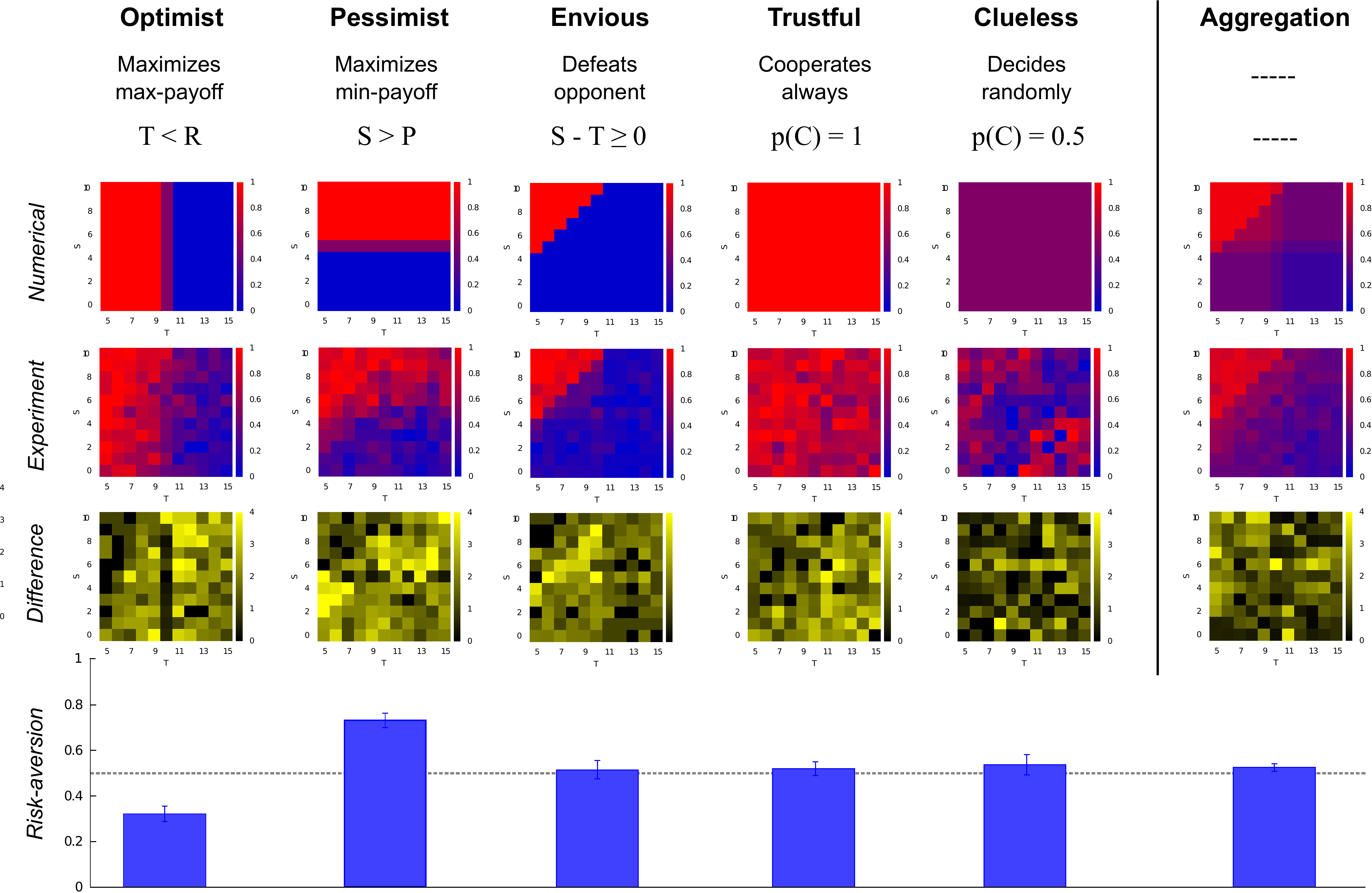}
\end{array}$
\end{center}
\caption{Difference between the experimental (second row) and numerical (or inferred, first row) behavioral heatmaps for each one of the phenotypes found by the $K$-means clustering algorithm, in units of SD. The difference between theory and experiment averaged over all $(T,S)$-plane is $1.91$ SD units for Envious, $1.85$ SD units for Optimist, $2.14$ SD units for Pessimist, $1.79$ SD units for Trustful, $1.12$ SD units for Undefined and $1.39$ SD units for the overall results in the Aggregation column.} \label{fig:differences_experiment_numerical}
\end{figure*}

\begin{figure}[h] 
\begin{center}$
\begin{array}{cc}
\includegraphics[width=0.95\textwidth]{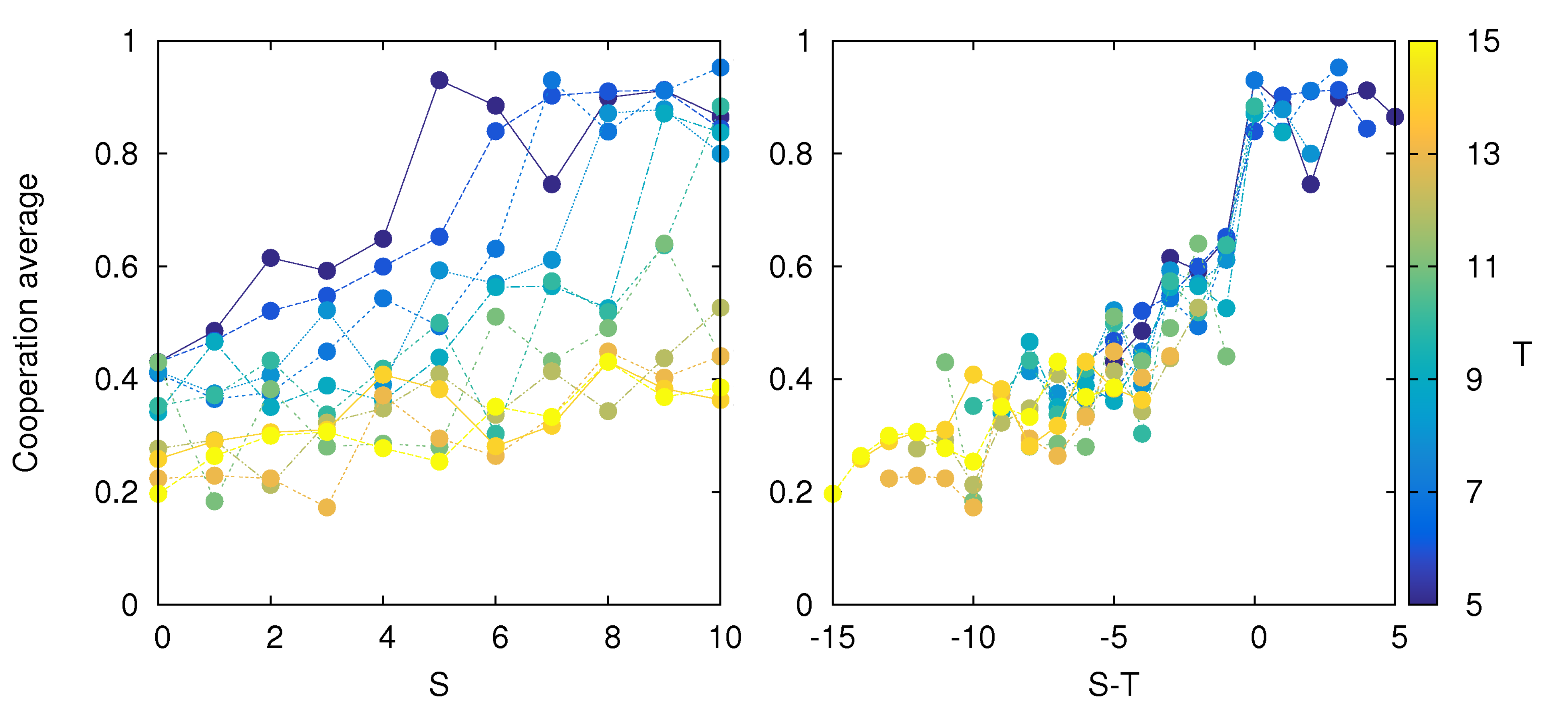}
\end{array}$
\end{center}
\caption{Average level of cooperation over all game actions and for different values of $T$ (in different colours). We observe disparate results when cooperation fraction is represented as a function of $S$ (left) but we find a nice collapse of all curves when cooperation level is expressed as a function $(S-T)$ (right).}
\label{fig:cooperation_vs_SminusT_all}
\end{figure}

\begin{figure}[h]   
\begin{center}$
\begin{array}{cc}
\includegraphics[width=0.85\textwidth]{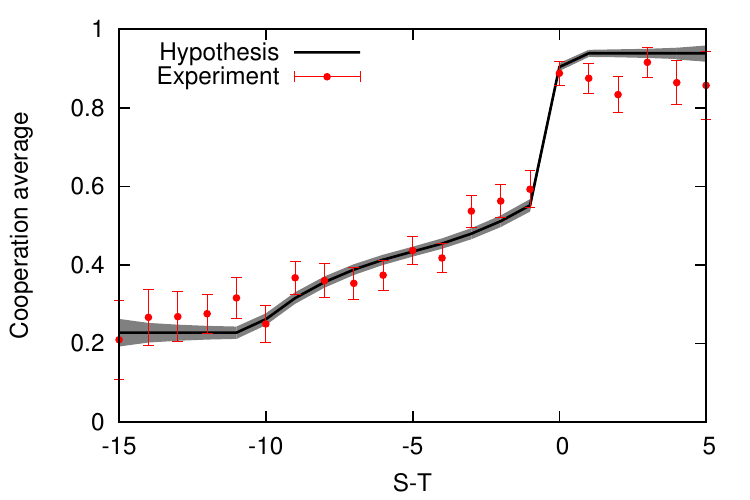}
\end{array}$
\end{center}
\caption{Average level of cooperation as a function of $(S-T)$ for both hypothesis and experiment. We consider the weight (number of decisions) in each cell when averaging over cells with same $(S-T)$. The error bars and the grey area represents a $95\%$ Confidence Interval for the experimental points and the recreated curve respectively.}\label{fig:S-T_final}
\end{figure}

\newpage

\end{document}